\documentclass[journal,draftcls,onecolumn,12pt,twoside]{IEEEtranTCOM}

\usepackage{graphicx}
\usepackage{epstopdf}
\usepackage{amsmath, amssymb, amsthm, amsfonts}
\usepackage[font=footnotesize]{subfig}
\usepackage{mathtools}
\usepackage{tikz}
\usetikzlibrary{shapes}
\usepackage{pgfplots}
\tikzset{>=latex}

\usepackage{cite}
\usepackage{multicol}

\renewcommand\IEEEkeywordsname{Index terms}

\newtheorem{theorem}{Theorem}

\newtheorem{corollary}{Corollary}

\newtheorem*{remark}{Remark}

\begin{document}
	
\bstctlcite{IEEEexample:BSTcontrol}

\title{The Cost of Delay in Status Updates and their Value: Non-linear Ageing}	

\author{Antzela~Kosta,~\IEEEmembership{Student Member,~IEEE,}
	Nikolaos~Pappas,~\IEEEmembership{Member,~IEEE,}
	Anthony~Ephremides,~\IEEEmembership{Life Fellow,~IEEE,}
	and~Vangelis~Angelakis,~\IEEEmembership{Senior Member,~IEEE}% <-this % stops a space
	\thanks{This work extends the preliminary study in \cite{Kosta17_ISIT}. 
		
	A. Kosta, N. Pappas, A. Ephremides, and V. Angelakis are with the Department of Science and Technology, Link{\"o}ping University, Norrk{\"o}ping SE-60174, Sweden (email: antzela.kosta@liu.se, nikolaos.pappas@liu.se, vangelis.angelakis@liu.se). 
	A. Ephremides is also with the Department of Electrical and Computer Engineering and the Institute for System Research, University of Maryland, College Park, MD 20740, USA (email: etony@ece.umd.edu).
		
	The research leading to these results has been partially funded by the
	European Union's Horizon 2020 research and innovation programme under the Marie Sklodowska-Curie Grant Agreement No. 642743 (WiVi-2020). 
	In addition, this work was supported in part by ELLIIT and the Center for Industrial Information Technology (CENIIT).
	The work of Anthony Ephremides is supported by the U.S. Office of Naval Research under Grant ONR 5-280542, the U.S. National Science Foundation under Grants CIF 5-243150, Nets 5-245770, and CIF 5-231912, and
	the Swedish Research Council (VR).}}% 

\maketitle

\vspace{-10mm}
\begin{abstract}
We consider a status update communication system consisting of a source-destination link.
A stochastic process is observed at the source, where samples are extracted at random time instances, and delivered to the destination, thus, providing status updates for the source.
In this paper, we expand the concept of information ageing by introducing the \emph{cost of update delay} (CoUD) metric to characterize the cost of having stale information at the destination.
The CoUD captures the freshness of the information at the destination and can be used to reflect the information structure of the source.
Moreover, we introduce the \emph{value of information of update} (VoIU) metric that captures the reduction of CoUD upon reception of an update.
Using the CoUD, its by-product metric called \emph{peak cost of update delay} (PCoUD), and the VoIU, we evaluate the performance of an M/M/1 system in various settings that consider exact expressions and bounds.
Our results indicate that the performance of CoUD differs depending on the cost assigned per time unit, however the optimal policy remains the same for linear ageing and varies for non-linear ageing.
When it comes to the VoIU the performance difference appears only when the cost increases non-linearly with time. 
The study illustrates the importance of the newly introduced variants of age, furthermore supported in the case of VoIU by its tractability.

\end{abstract}


\IEEEpeerreviewmaketitle

\section{Introduction}
A wide range of applications, from sensor networking to the stock market, requires the timely monitoring of a remote system. To quantify the freshness of such information, the concept of age of information (AoI) was introduced in \cite{Kaul12_INFOCOM,Song1990}. To define age consider that a monitored node generates status updates, timestamps them and transmits them over some network to a destination. Then, the age of the information the destination has for the source, or more simply AoI, is the time that elapsed from the generation of the last received status update. 
The notion of data freshness goes back to real-time databases studies \cite{Song1990,Segev1991,Adelberg1995_ACM,Cho2000_ACM}.
Keeping the average AoI small corresponds to having fresh information.

Part of AoI research has so far focused on the use of different queueing models through which the status updates may be processed.
The average age has been investigated in \cite{Kaul12_INFOCOM} for the M/M/1, D/M/1, and M/D/1 queues.
Minimizing AoI over the space of all update generation and service time distributions was analysed in \cite{Talak18_arXiv,Inoue18_arXiv}.
In \cite{Kam16} the authors investigate the performance of M/M/1, M/M/2, and M/M/$\infty$ cases, adding to the system model a more complex network feature. 
In their work packets travel over a network that may have out-of-order delivery, for example due to route diversity. 
Thus the notion of obsolete, non-informative, packets arises.
Multiple sources are studied in \cite{Sun18_INFOCOM,Najm18_INFOCOM,Yates19_transactions}, where the authors characterize how the serving facility can be shared among multiple update sources. 
In \cite{Costa16}, the notion of \emph{peak age of information} (PAoI) was introduced and characterized.
%Furthermore, a newly proposed metric related with AoI was presented in \cite{Costa14_ISIT}, called \emph{peak age of information} (PAoI).
In \cite{Modiano15_ISIT}, the authors consider the problem of optimizing the PAoI by controlling the arrival rate of update messages and derive properties of the optimal solution for the M/G/1 and M/G/1/1 models.

Controlling the messages' handling policy can increase the performance, starting from a simple last-generated-first-served (LGFS) service discipline \cite{Kaul12_CISS,Najm16_ISIT}, to more complicated packet management that discards non-informative packets \cite{Costa16,Pappas15_ICC,Kosta19_ISIT,Kosta19_JCN,Soysal19_arXiv}.
In \cite{Kam2018_transactions} the authors introduce a packet deadline as a control mechanism and study its impact on the average age of an M/M/1/2 system.
In \cite{Baknina18_ISIT}, the authors consider the scenario where the timings of the status updates also carry an independent message and study the tradeoff between the achievable message rate and the achievable average AoI.
%In \cite{Chen16_ISIT} on the other hand, the authors take into consideration packet delivery errors, i.e., update packets can get lost during a transmission to their destination.
In \cite{Bedewy16_ISIT} the authors consider multiple servers where each server can be viewed as a wireless link.
They prove that a preemptive LGFS service simultaneous optimizes the age, throughput, and delay performance in infinite buffer queueing systems.
In \cite{Bedewy17_ISIT,Talak17_Allerton,Yates18_arXivSHS,Yates18_INFOCOM} the minimization of age is done over general multihop networks.
A wireless network with heterogeneous traffic is considered in \cite{Kosta18_GLOBECOM,Stamatakis19_GLOBECOM,Buyukates19_arXiv}.
Another control policy is to assume that the source is monitoring the network servers' idle/busy state and is able to generate status updates at any time, as in \cite{Yates15_ISIT,Sun16_INFOCOM,Soleymani18_arXiv,Stamatakis19_arXiv_GLOBECOM}.
Aging control policies in hybrid networks are studied in \cite{Altman19_Mobihoc,Azouzi12_ICST}.

Apart from system considerations such as the the arrival process, the queueing model, and the service process, the characteristics of the source observed process can play an important role in the chosen frequency of status update transmissions. 
In fact, depending on the context, age can be modified to migrate to an effective age, different for each application.
Non-linear utility functions in the context of dynamic content distribution were considered in the past in \cite{Cho2003_ACM,Even2007_ACM,Heinrich2009_JDIQ,Ioannidis09_INFOCOM,Razniewski2016_ACM}, however the queueing aspect along with freshness is not captured.
In \cite{Sun2017_transactions,Sun18_JCN}, a so called age penalty/utility function was employed to describe the level of dissatisfaction for having aged status updates at the destination.
In \cite{Sun18_SPAWC}, the authors use the mutual information between the real-time source value and the delivered samples at the receiver to quantify the freshness of the information contained in the delivered samples.
In \cite{Poojary17_ITW}, an incremental update scheme for real time status updates which exploits temporal correlation between consecutive messages is considered.
In \cite{Sun17_ISIT} it was proven that for a Wiener process sampling that minimizes AoI is not optimal for prediction.
In \cite{Kam18_SPAWC,Kam18_INFOCOM}, the authors propose various effective age metrics to minimize the estimation/prediction error.
%Exploring the time correlation of a source has been studied in 
Exploring the correlation among multiple sources has been studied in \cite{He18_INFOCOM,Hribar17_GLOBECOM}.
For a more broad view of the works on AoI we direct the reader to our summary article in \cite{NET-060}.

\subsection{Contribution}

The need to go beyond AoI to characterize the level of \lq\lq{}dissatisfaction\rq\rq{} for data staleness has been first reflected in \cite{Sun16_INFOCOM}, where an age penalty was introduced. 
However, there is also a need to capture not only the application demand for fresh data, but also the source signal properties as well.
Pushing forward, we introduce the \emph{cost of update delay} (CoUD) metric for three sample case functions that can be easily tuned through a parameter.
For each case, we derive the time average cost for an M/M/1 model with a first-come-first-served (FCFS) queue discipline.
In addition, we derive upper bounds to the average CoUD that may be used for further system design and optimization, and establish their association with the by-product of CoUD, called, \emph{peak cost of update delay} (PCoUD).
Although in \cite{Sun16_INFOCOM} penalty functions are said to be determined by the application, we go further and associate the cost of staleness with the statistics of the source.

Before defining this association, we first need to elaborate on the requirement of small AoI.
Why are we interested in small AoI?
Consider that we are observing a system at time instant $t$.
However, the most recent value of the observed process available is the one that had arrived at $t-\Delta$, for some random $\Delta$.
Now assume that the destination node wants to estimate the information at time $t$.
If the samples at $t$ and $t-\Delta$ are independent, the knowledge of $t-\Delta$ is not useful for the estimation and age simply indicates delay.
However, if the samples at $t$ and $t-\Delta$ are dependent, then the value of $\Delta$ will affect the accuracy of the estimation.
A smaller $\Delta$ can lead to a more accurate estimation.
Our work is a first step towards exploring this potential usage of AoI.

Next, we introduce a novel metric called \emph{value of information of update} (VoIU) to capture the degree of importance of the information received at the destination.
A newly received update reduces the uncertainty of the destination about the current value of the observed stochastic process, and VoIU captures that reduction that is directly related to the time elapsed since the last update reception.
Following this approach, we take into consideration not only the probability of a reception event, but also the impact of the event on our knowledge of the evolution of the process.

Small CoUD corresponds to timely information while VoIU represents the impact of the received information in reducing the CoUD.
Therefore, in a communication system it would be highly desirable to minimize the average CoUD, and at the same time maximize the average VoIU.
To this end we derive the average VoIU for the M/M/1 queue and discuss how the optimal server utilization with respect to VoIU can be used in relation with the CoUD average analysis.

\section{System Model and Definitions}
\label{sec:model}
We consider a system in which a source generates status updates in the form of packets at random intervals.
These packets contain the status update information and a timestamp of their generation time.
The generated packets are queued and then transmitted over a link according to an FCFS queue discipline, to reach a remote destination.

To assess the freshness of the randomly generated updates, Kaul et al. \cite{Kaul12_INFOCOM} defined \emph{age of information}, $\Delta(t) = t - u(t)$, to be the difference of the current time instant and the timestamp of the last received update. 
In this paper, we expand the notion of \emph{age} by defining \emph{cost of update delay} (CoUD)
\begin{equation}
C(t) = f_s(t - u(t)),
\label{eq:cost_t}
\end{equation} 
to be a stochastic process that increases as a function of time between received updates. 
We introduce a non-negative, monotonically increasing category of functions $f_s(t)$, having $f_s(0)=0$, to represent the evolution of the cost of update delay according to the characteristics of the source data.
In the absence of status updates at the destination, this staleness metric increases as a function of time, while upon reception of a new status update, the \emph{cost} drops to a smaller value that is equal to the delay of that update.
Different CoUD functions enable us to capture the freshness of the process under observation, and implicitly the autocorrelation structure of the source signal.
This has further implications that go beyond the scope of this work.

\begin{figure}[t]\centering
	\centering
	\begin{tikzpicture}[scale=0.9]
% horizontal axis
\draw[->] (0,0) -- (8.2,0) node[anchor=north] {$t$};
% vertical axis
\draw[->] (0,0) -- (0,3.5) node[anchor=east] {$C(t)$};

%\draw	(-0.5,2) node[rotate=90] {Age};
\draw	(-0.3,0.25) node[anchor=south] {$C_0$};

%shadow area
 \draw[fill=gray!10] (1.4,0) -- (4.4,3) -- (4.4,2.4)-- (2,0);

% labels
\draw	(-0.5,0) node[anchor=north] {$t_0$}
           (0.2,0) node[anchor=north] {$t_1$}
		    (1.4,0) node[anchor=north] {$t_2$}
		    (2,0) node[anchor=north] {$t_3$}
		    (3.8,0) node[anchor=north] {$t_4$}
		    (6,0) node[anchor=north] {$t_{n-1}$}
		    (6.8,0) node[anchor=north] {$t_n$};
		    
\draw[->,>=stealth]    (1,0) -- (1,-0.4) node[anchor=south,below] {$t'_1$};
\draw[->,>=stealth]  (2.5,0) -- (2.5,-0.4) node[anchor=south,below] {$t'_2$};
\draw[->,>=stealth]  (4.4,0) -- (4.4,-0.4) node[anchor=south,below] {$t'_3$};
\draw[->,>=stealth]   (7.4,0) -- (7.4,-0.4) node[anchor=south,below] {$t'_n$};
		    	    		 
\draw	(0.55,0.7) node{{\scriptsize $Q_1$}}
		    (1.6,0.75) node{{\scriptsize $Q_2$}};
           %(3.1,2.5) node{{\scriptsize $Q_3$}};
\draw   (3.7,0.75) node{{\scriptsize $Q_4$}}
           (7,0.7) node{{\scriptsize $Q_n$}}
           (7.23,0.2) node{{\scriptsize $\tilde{Q}$}};
           
\draw[<-] (3.3,1.5) to [out=95,in=250] (3.3,2.4) node [above] {{\scriptsize $Q_3$}};           
           
\draw [thick](0.2,-1.2) -- (1.4,-1.2) node[pos=.5,sloped,below] {$Y_2$} ;
\draw[thick]  (0.2,-1.3) -- (0.2,-1.1); 
\draw [thick](1.4,-1.2) -- (2.5,-1.2) node[pos=.5,sloped,below] {$T_2$} ;
\draw[thick]  (1.4,-1.3) -- (1.4,-1.1) 
                    (2.5,-1.3) -- (2.5,-1.1);
                    
\draw [thick](6,-1.2) -- (6.8,-1.2) node[pos=.5,sloped,below] {$Y_n$} ;
\draw[thick]  (6,-1.3) -- (6,-1.1); 
\draw [thick](6.8,-1.2) -- (7.4,-1.2) node[pos=.5,sloped,below] {$T_n$} ;
\draw[thick]  (6.8,-1.3) -- (6.8,-1.1) 
                    (7.4,-1.3) -- (7.4,-1.1);
 
%Q1
\draw[thick] (0,0.5) -- (1,1.5) -- (1,0.8);
%sawtooth
\draw[thick] (0.2,0) -- (2.5,2.3) -- (2.5,1.1) -- (4.4,3) -- (4.4,2.4) -- (4.7,2.7);
%Qn
\draw[thick] (6.4,0.5) -- (7.4,1.5) -- (7.4,0.6);

%colored part
\draw[thick,red] (1,0.8) -- (1,1.5);
\draw[thick,red] (2.5,1.1) -- (2.5,2.3);
\draw[thick,red] (4.4,2.4) -- (4.4,3);
\draw[thick,red] (7.4,0.6) -- (7.4,1.5);

%vertical lines
\draw[dotted] (1,0) -- (1,0.8);
\draw[dotted] (2.5,0) -- (2.5,1.1);
\draw[dotted] (4.4,0) -- (4.4,2.4);
\draw[dotted] (7.4,0) -- (7.4,0.6);
%diagonal lines
\draw[dotted] (-0.5,0) -- (0,0.5);
\draw[dotted] (1.4,0) -- (2.5,1.1);
\draw[dotted] (2,0) -- (4.7,2.7);
\draw[dotted] (3.8,0) -- (5,1.2);
\draw[dotted] (5.9,0) -- (6.4,0.5);
\draw[dotted] (6.8,0) -- (7.4,0.6);

%infinity symbol
\draw[thick]  (5.2,-0.15) -- (5.2,0.15) 
                    (5.3,-0.15) -- (5.3,0.15);
\draw[white, fill=white!50] (5.21,-0.2) -- (5.21,0.2) -- (5.29,0.2) -- (5.29,-0.2) ;                   

\draw [decorate,decoration={brace,amplitude=5pt,mirror,raise=2pt},yshift=0pt] (1,0.8) -- (1,1.55) node [black,midway,yshift=0.3cm,xshift=0.35cm] {\scriptsize $D_1$};
\draw [decorate,decoration={brace,amplitude=5pt,mirror,raise=2pt},yshift=0pt] (2.47,1.1) -- (2.47,2.33) node [black,midway,yshift=0.1cm,xshift=0.39cm] {\scriptsize $D_2$};
\draw [decorate,decoration={brace,amplitude=4pt,mirror,raise=2pt},yshift=0pt] (4.36,2.4) -- (4.36,3) node [black,midway,yshift=0.1cm,xshift=0.45cm] {\scriptsize $D_3$};
\draw [decorate,decoration={brace,amplitude=5pt,mirror,raise=2pt},yshift=0pt] (7.4,0.6) -- (7.4,1.5) node [black,midway,xshift=0.5cm] {\scriptsize $D_n$};
\end{tikzpicture}
	\vspace{-3mm}
	\caption{Example of the linear CoUD evolution over time.}
	\label{fig:linear_value}
\end{figure}
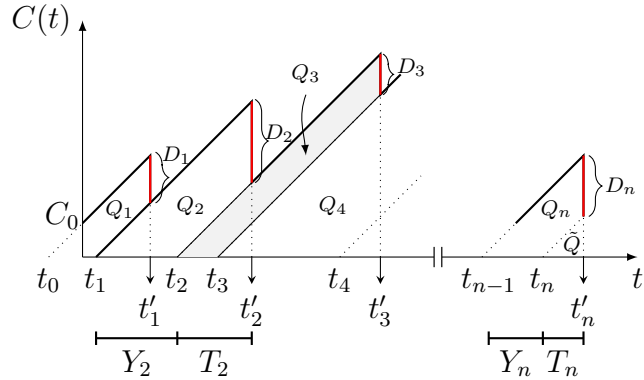

Update $i$ is generated at time $t_i$ and is received by the destination at time $t'_i$.
The cost of information absence at the destination increases as a function $f_s(t)$ of time.
Note that age as coined by Kaul is a special cost case, where the cost is counted in time units, as shown in Fig.~\ref{fig:linear_value}. 
In this paper we consider that the cost can take any form of a ``payment'' function that can also assign to it any relevant unit.

The $i$th interarrival time $Y_i = t_i - t_{i-1}$ is the time elapsed between the generation of update $i$  and the previous update generation and is a random variable.
Moreover,  $T_i = t'_i  - t_i$ is the system time of update $i$ corresponding to the sum of the waiting time at the queue and the service time.
Note that the random variables $Y_i$ and $T_i$ are real system time measures and are independent of the way we choose to calculate the cost of update delay i.e., of $f_s(t)$.

The value of CoUD achieved immediately before receiving the $i$th update is called, in analogy to  the peak age \cite{Costa14_ISIT}, \emph{peak cost of update delay} (PCoUD), and is defined as
\begin{equation}
A_i = f_s(t'_i -t_{i-1}).
\label{eq:A_i}
\end{equation}

At time $t'_i$, the cost $C(t'_i)$ is reset to $f_s(t'_i  - t_i)$ and we introduce the \emph{value of information of update} (VoIU) $i$ as
\begin{equation}
V_i = \frac{f_s(t'_i -t_{i-1}) - f_s(t'_i - t_i)}{f_s(t'_i - t_{i-1})},
\label{eq:V_i}
\end{equation} 
to measure the degree of importance of the status update received at the destination.
Intuitively, this metric depends on two system parameters at the time of observation: (i) the cost of update delay at the destination (ii) the time that the received update was generated. 
This can be easily shown to be similarly expressed as a dependence on: (i) the interarrival time of the last two packets received (ii) the current reception time. 

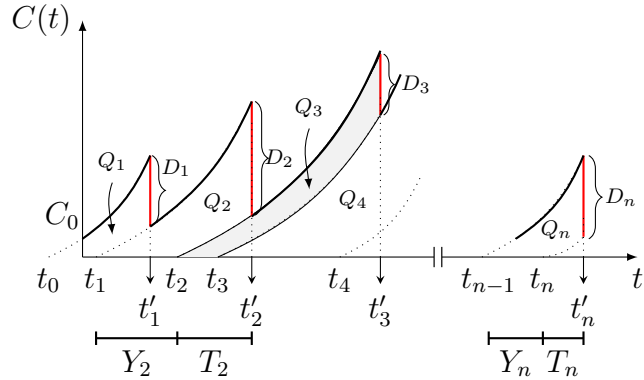
\begin{figure}[t]\centering
	\centering
	\begin{tikzpicture}[scale=0.9]
% horizontal axis
\draw[->] (0,0) -- (8.2,0) node[anchor=north] {$t$};
% vertical axis
\draw[->] (0,0) -- (0,3.5) node[anchor=east] {$C(t)$};
%\draw	(-0.5,2) node[rotate=90] {Age};
\draw	(-0.3,0.25) node[anchor=south] {$C_0$};

%shadow area
 \draw[fill=gray!10] (1.4,0) to[bend right=20] (4.4,3) -- (4.4,2.1) to[bend left=17] (2,0);
% labels
\draw	(-0.5,0) node[anchor=north] {$t_0$}
           (0.2,0) node[anchor=north] {$t_1$}
		    (1.4,0) node[anchor=north] {$t_2$}
		    (2,0) node[anchor=north] {$t_3$}
		    (3.8,0) node[anchor=north] {$t_4$}
		    (6,0) node[anchor=north] {$t_{n-1}$}
		    (6.8,0) node[anchor=north] {$t_n$};
%receptions		    
\draw[->,>=stealth]    (1,0) -- (1,-0.4) node[anchor=south,below] {$t'_1$};
\draw[->,>=stealth]  (2.5,0) -- (2.5,-0.4) node[anchor=south,below] {$t'_2$};
\draw[->,>=stealth]  (4.4,0) -- (4.4,-0.4) node[anchor=south,below] {$t'_3$};
\draw[->,>=stealth]   (7.4,0) -- (7.4,-0.4) node[anchor=south,below] {$t'_n$};
		    	    		 
\draw	%(0.4,1.2) node{{\scriptsize $Q_1$}}
		    (2,0.8) node{{\scriptsize $Q_2$}};
%\draw   (3.3,1.8) node{{\scriptsize $Q_3$}};
\draw   (4.0,0.85) node{{\scriptsize $Q_4$}}
           %(3.6,0.8) node{{\scriptsize $Q_4$}}
           (7,0.4) node{{\scriptsize $Q_n$}};
           
\draw[<-] (0.44,0.33) to [out=95,in=250] (0.44,1.1) node [above] {{\scriptsize $Q_1$}};         
                      
\draw[<-] (3.35,1.0) to [out=95,in=250] (3.35,1.9) node [above] {{\scriptsize $Q_3$}};         

%\draw[<-] (7.23,0.08) to [in=95,out=250] (8.15,0.1) node [above] {{\scriptsize $\tilde{Q}$}};         
           
\draw [thick](0.2,-1.2) -- (1.4,-1.2) node[pos=.5,sloped,below] {$Y_2$} ;
\draw[thick]  (0.2,-1.3) -- (0.2,-1.1); 
\draw [thick](1.4,-1.2) -- (2.5,-1.2) node[pos=.5,sloped,below] {$T_2$} ;
\draw[thick]  (1.4,-1.3) -- (1.4,-1.1) 
                    (2.5,-1.3) -- (2.5,-1.1);
                    
\draw [thick](6,-1.2) -- (6.8,-1.2) node[pos=.5,sloped,below] {$Y_n$} ;
\draw[thick]  (6,-1.3) -- (6,-1.1); 
\draw [thick](6.8,-1.2) -- (7.4,-1.2) node[pos=.5,sloped,below] {$T_n$} ;
\draw[thick]  (6.8,-1.3) -- (6.8,-1.1) 
                    (7.4,-1.3) -- (7.4,-1.1);
 
%main vertical lines
\draw[thick,red]  (1,1.5) -- (1,0.45);
\draw[thick,red] (2.5,2.3) -- (2.5,0.6);
\draw[thick,red] (4.4,3) -- (4.4,2.1);
\draw[thick,red] (7.4,1.5) -- (7.4,0.3);

%curves (default=30)
\draw[dotted]  (-0.5,0) to[bend right=20] (1,1.5); %Q1
\draw[thick]  (0,0.27) to[bend right=15] (1,1.5); %Q1

\draw[dotted]  (0.2,0) to[bend right=20] (2.5,2.3); %Q2
\draw[thick]  (1,0.45) to[bend right=15] (2.5,2.3); %Q2

\draw[dotted]  (1.4,0) to[bend right=20] (4.4,3); %Q3
\draw[thick]  (2.5,0.6) to[bend right=15] (4.4,3.05); %Q3

\draw[dotted]  (2,0) to[bend right=20] (4.7,2.7); %Q4
\draw[thick]  (4.4,2.1) to[bend right=3] (4.7,2.7); %Q4

\draw[dotted]  (5.9,0) to[bend right=20] (7.4,1.5); %Qn
\draw[thick]  (6.4,0.27) to[bend right=17] (7.4,1.5); %Qn

%vertical lines
\draw[dotted] (1,0) -- (1,0.5);
\draw[dotted] (2.5,0) -- (2.5,2);
\draw[dotted] (4.4,0) -- (4.4,2.4);
\draw[dotted] (7.4,0) -- (7.4,0.6);

%%diagonal lines
\draw[dotted] (3.8,0) to[bend right=20] (5,1.2);
\draw[dotted] (6.8,0) to[bend right=20] (7.4,0.3);

%infinity symbol
\draw[thick]  (5.2,-0.15) -- (5.2,0.15) 
                    (5.3,-0.15) -- (5.3,0.15);
\draw[white, fill=white!50] (5.21,-0.2) -- (5.21,0.2) -- (5.29,0.2) -- (5.29,-0.2) ;                   

\draw [decorate,decoration={brace,amplitude=5pt,mirror,raise=2pt},yshift=0pt] (0.95,0.47) -- (0.95,1.5) node [black,midway,yshift=0.3cm,xshift=0.39cm] {\scriptsize $D_1$};
\draw [decorate,decoration={brace,amplitude=5pt,mirror,raise=2pt},yshift=0pt] (2.47,0.62) -- (2.47,2.3) node [black,midway,xshift=0.39cm] {\scriptsize $D_2$};
\draw [decorate,decoration={brace,amplitude=4pt,mirror,raise=2pt},yshift=0pt] (4.37,2.1) -- (4.37,3) node [black,midway,xshift=0.5cm] {\scriptsize $D_3$};
\draw [decorate,decoration={brace,amplitude=5pt,mirror,raise=2pt},yshift=0pt] (7.4,0.27) -- (7.4,1.5) node [black,midway,xshift=0.5cm] {\scriptsize $D_n$};              
\end{tikzpicture}
	\vspace{-3mm}
	\caption{Example of the exponential CoUD evolution over time.}
	\label{fig:exponential_value}
\end{figure}

\subsection{CoUD Properties}

To explore a wide array of potential uses of the notion of \emph{cost}, we investigate three sample cases for the $f_s(\cdot)$ function
%\begin{align}
%f_s(t) =
%\begin{cases} 
%\alpha t \\ 
%e^{\alpha  t}  -1\\
%\log(\alpha  t +1)
%\end{cases} 
%\end{align}
\begin{equation}
	f_s(t) = \alpha t,
	\label{eq:fs_lin}
\end{equation}
\begin{equation}	
    f_s(t) = e^{\alpha  t}  -1,
    \label{eq:fs_exp}
\end{equation}
\begin{equation}
	 f_s(t) = \log(\alpha  t +1),	
	 \label{eq:fs_log}
\end{equation}
for $\alpha > 0$.
As mentioned earlier, we can not fully leverage CoUD if we do not assume that the samples of the observed stochastic process are dependent.
With this in mind, we propose the selection of the $f_s(\cdot)$ function of CoUD according to the autocorrelation of the process. 
More specifically, if the autocorrelation is small, we suggest the exponential function, in order to penalize the increase of system time between updates, which would significantly affect the reconstruction potential of the process.
If the autocorrelation is large, the logarithmic function is more appropriate. 
For intermediate values the linear case can be a reasonable choice.

\begin{figure}[t]\centering
	\centering
	\begin{tikzpicture}[scale=0.9]
% horizontal axis
\draw[->] (0,0) -- (8.2,0) node[anchor=north] {$t$};
% vertical axis
\draw[->] (0,0) -- (0,3.5) node[anchor=east] {$C(t)$};
%\draw	(-0.5,2) node[rotate=90] {Age};
\draw	(-0.3,0.25) node[anchor=south] {$C_0$};

%shadow area
 \draw[fill=gray!10] (1.4,0) to[bend left=20] (4.4,3) -- (4.4,2.55) to[bend right=19] (2,0);
% labels
\draw	(-0.5,0) node[anchor=north] {$t_0$}
           (0.2,0) node[anchor=north] {$t_1$}
		    (1.4,0) node[anchor=north] {$t_2$}
		    (2,0) node[anchor=north] {$t_3$}
		    (3.8,0) node[anchor=north] {$t_4$}
		    (6,0) node[anchor=north] {$t_{n-1}$}
		    (6.8,0) node[anchor=north] {$t_n$};
%receptions		    
\draw[->,>=stealth]    (1,0) -- (1,-0.4) node[anchor=south,below] {$t'_1$};
\draw[->,>=stealth]  (2.5,0) -- (2.5,-0.4) node[anchor=south,below] {$t'_2$};
\draw[->,>=stealth]  (4.4,0) -- (4.4,-0.4) node[anchor=south,below] {$t'_3$};
\draw[->,>=stealth]   (7.4,0) -- (7.4,-0.4) node[anchor=south,below] {$t'_n$};
		    	    		 
\draw	%(0.4,1.7) node{{\scriptsize $Q_1$}}
		    (1.4,0.9) node{{\scriptsize $Q_2$}};
%\draw   (3.3,2.7) node{{\scriptsize $Q_3$}};
\draw   (4.0,1.3) node{{\scriptsize $Q_4$}}
           %(3.6,0.8) node{{\scriptsize $Q_4$}}
           (6.8,0.5) node{{\scriptsize $Q_n$}};
           %(7.23,0.15) node{{\scriptsize $\tilde{Q}$}}
           
\draw[<-] (0.44,0.75) to [out=95,in=250] (0.44,1.65) node [above] {{\scriptsize $Q_1$}};    
           
\draw[<-] (3.3,2.05) to [out=95,in=250] (3.3,2.95) node [above] {{\scriptsize $Q_3$}};            
           
\draw [thick](0.2,-1.2) -- (1.4,-1.2) node[pos=.5,sloped,below] {$Y_2$} ;
\draw[thick]  (0.2,-1.3) -- (0.2,-1.1); 
\draw [thick](1.4,-1.2) -- (2.5,-1.2) node[pos=.5,sloped,below] {$T_2$} ;
\draw[thick]  (1.4,-1.3) -- (1.4,-1.1) 
                    (2.5,-1.3) -- (2.5,-1.1);
                    
\draw [thick](6,-1.2) -- (6.8,-1.2) node[pos=.5,sloped,below] {$Y_n$} ;
\draw[thick]  (6,-1.3) -- (6,-1.1); 
\draw [thick](6.8,-1.2) -- (7.4,-1.2) node[pos=.5,sloped,below] {$T_n$} ;
\draw[thick]  (6.8,-1.3) -- (6.8,-1.1) 
                    (7.4,-1.3) -- (7.4,-1.1);
 
%main vertical lines
\draw[thick,red]  (1,1.48) -- (1,1.25);
\draw[thick,red] (2.5,2.3) -- (2.5,1.7);
\draw[thick,red] (4.4,3) -- (4.4,2.55);
\draw[thick,red] (7.4,1.5) -- (7.4,0.4);

%curves (default=30)
\draw[dotted]  (-0.5,0) to[bend left=20] (1,1.5); %Q1
\draw[thick]  (0,0.8) to[bend left=14] (1,1.48); %Q1

\draw[dotted]  (0.2,0) to[bend left=20] (2.5,2.3); %Q2
\draw[thick]  (1,1.25) to[bend left=11] (2.5,2.3); %Q2

\draw[dotted]  (1.4,0) to[bend left=20] (4.4,3); %Q3
\draw[thick]  (2.5,1.7) to[bend left=10] (4.4,3.0); %Q3

\draw[dotted]  (2,0) to[bend left=20] (4.7,2.7); %Q4
\draw[thick]  (4.4,2.55) to[bend left=3] (4.7,2.7); %Q4

\draw[dotted]  (5.9,0) to[bend left=20] (7.4,1.5); %Qn
\draw[thick]  (6.4,0.8) to[bend left=11] (7.4,1.5); %Qn

%vertical lines
\draw[dotted] (1,0) -- (1,1.25);
\draw[dotted] (2.5,0) -- (2.5,2);
\draw[dotted] (4.4,0) -- (4.4,2.4);
\draw[dotted] (7.4,0) -- (7.4,0.6);

%%diagonal lines
\draw[dotted] (3.8,0) to[bend left=20] (5,1.2);
\draw[dotted] (6.8,0) to[bend left=20] (7.4,0.4);

%infinity symbol
\draw[thick]  (5.2,-0.15) -- (5.2,0.15) 
                    (5.3,-0.15) -- (5.3,0.15);
\draw[white, fill=white!50] (5.21,-0.2) -- (5.21,0.2) -- (5.29,0.2) -- (5.29,-0.2) ;                   

\draw [decorate,decoration={brace,amplitude=2pt,mirror,raise=2pt},yshift=0.5pt] (0.95,1.26) -- (0.95,1.5) node [black,midway,yshift=0.4cm,xshift=0.29cm] {\scriptsize $D_1$};
\draw [decorate,decoration={brace,amplitude=4pt,mirror,raise=2pt},yshift=0pt] (2.47,1.75) -- (2.47,2.3) node [black,midway,yshift=0.25cm,xshift=0.32cm] {\scriptsize $D_2$};
\draw [decorate,decoration={brace,amplitude=3pt,mirror,raise=2pt},yshift=0pt] (4.37,2.57) -- (4.37,3) node [black,midway,yshift=0.05cm,xshift=0.4cm] {\scriptsize $D_3$};
\draw [decorate,decoration={brace,amplitude=5pt,mirror,raise=2pt},yshift=0pt] (7.4,0.4) -- (7.4,1.5) node [black,midway,xshift=0.5cm] {\scriptsize $D_n$};              
\end{tikzpicture}
	\vspace{-3mm}
	\caption{Example of the logarithmic CoUD evolution over time.}
	\label{fig:logarithmic_value}
\end{figure}
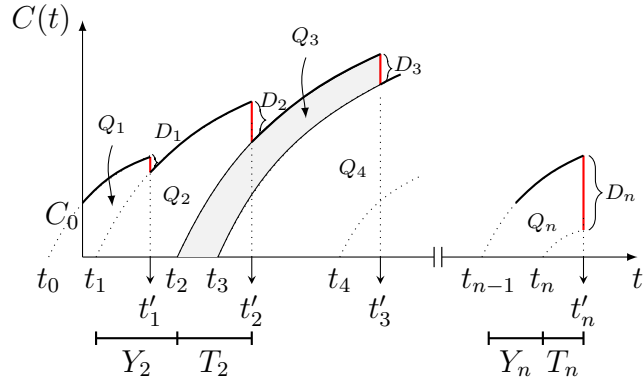

\begin{remark}\label{remark1}
Observe that it would not make sense to have the exponential cost increase beyond the maximum value of the prediction error.
In the special case that the observed process is stationary this is constant and equal to the variance of the process as the cost goes to infinity.
Recall that one of the motivations of the works on AoI is the remote estimation and reconstruction of the source signal.
\end{remark}

The autocorrelation $R(t_1,t_2)=\mathbb{E}[x(t_1) x^*(t_2)]$ of a stochastic process is a positive definite function, that is $\sum_{i,j} \beta_i \beta_j^* R(t_i,t_j) >0$, for any $\beta_i$ and $\beta_j$.
Tuning the parameter $\alpha$ in \eqref{eq:fs_lin}-\eqref{eq:fs_log} properly enables us to associate with accuracy the right $f_s(\cdot)$ function to a corresponding autocorrelation.
%The reason is that when the process changes faster we want a concave increase convex
Next, we focus on VoIU and analyze it for each case of $f_s(t)$ separately. 

\section{Value of Information of Update Analysis}

The value of information of update is a bounded fraction that takes values in the real interval $\left[0,1\right] $, with 0 representing the minimum benefit of an update and 1 the maximum.
%\begin{lemma}
	In a system where status updates are instantaneously available from the source to the destination, VoIU is given by:
	\vspace{-1mm}
	\begin{equation}
	V_i = \lim_{t'_i \to t_i} \frac{f_s(t'_i -t_{i-1}) - f_s(t'_i - t_i)}{f_s(t'_i - t_{i-1})} = 1.
	\label{eq:property}
	\end{equation} 
%\end{lemma}
%The previous can be obtained from \eqref{eq:V_i} by taking the limit.
The interpretation of this property is that in the extreme case when the system time is insignificant and a packet reaches the destination as soon as it is generated, we assign to the VoIU metric the maximum value reflecting that the reception occurs without value loss.

We first derive useful results for the general case without considering specific queueing models.
For the first case, $f_s(t)=\alpha t$, expression (\ref{eq:V_i}) yields  
\begin{equation}
V_{P,i} = \frac{Y_i}{Y_i + T_i}.
\label{eq:Vi_at}
\end{equation}
Note that for $\alpha = 1$ the cost of update delay corresponds to the timeliness of each status update arriving and is the so called \emph{age of information}.
The cost reductions $\{ D_1, \dots , D_n \}$, depicted in Fig.~\ref{fig:linear_value}, correspond to the interarrival times $\{ Y_1, \dots , Y_n \}$, and also the limits,
%\begin{equation}
$\lim_{Y_i \to +\infty} V_{P,i} = 1,$
%\label{eq:lim_Vi_at}
%\end{equation}
and
%\begin{equation}
$\lim_{T_i \to +\infty} V_{P,i} = 0,$
%\label{eq:lim_Vi_at2}
%\end{equation}
agree with the definition.
Next, for $f_s(t)=e^{\alpha  t} -1$, shown in Fig.~\ref{fig:exponential_value}, the definition of VoIU is 
\begin{equation}
V_{E,i} = \frac{e^{\alpha (Y_i + T_i)} - e^{\alpha T_i}}{e^{\alpha (Y_i + T_i)}-1},
\label{eq:Vi_expat}
\end{equation} 
%and the corresponding limits are $\lim_{Y_i \to +\infty} V_{E,i} = 1$, $\lim_{T_i \to +\infty} V_{E,i} = 1-e^{-\alpha Y_i}$.
and the corresponding limits are 
%\begin{equation}
$\lim_{Y_i \to +\infty} V_{E,i} = 1,$
%\label{eq:lim_Vi_exp_at}
%\end{equation}
and
%\begin{equation}
$\lim_{T_i \to +\infty} V_{E,i} = 1-e^{-\alpha Y_i},$
%\label{eq:lim_Vi_exp_at2}
%\end{equation}
where $\lim_{Y_i \to 0} (1-e^{-\alpha Y_i}) = 0$, and $\lim_{Y_i \to +\infty} (1-e^{-\alpha Y_i}) = 1$.
At last, for the case $f_s(t)=\log(\alpha t +1)$, depicted in  Fig.~\ref{fig:logarithmic_value}, we obtain
\begin{equation}
V_{L,i} =\frac{\log(\alpha (Y_i + T_i) +1) - \log(\alpha T_i+1)}{\log(\alpha (Y_i + T_i) +1)}, 
\label{eq:Vi_logat}
\end{equation}
and the corresponding limits are 
%\begin{equation}
$\lim_{Y_i \to +\infty} V_{L,i} = 1,$
%\label{eq:lim_Vi_log_at}
%\end{equation}
%\begin{equation}
$\lim_{T_i \to +\infty} V_{L,i} = 0.$
%\label{eq:lim_Vi_log_at2}
%\end{equation}

The results above can be interpreted as follows. 
\emph{As the interarrival time of the received packets becomes large, the value of information of the updates takes its maximum value, underlining the importance to have a new update as soon as possible.
On the other hand, when the system time gets significantly large, we expect that the received update is not as timely as we would prefer in order to maintain the freshness of the system, hence we assign to the VoIU metric the minimum value.}

Suppose that our interval of observation is $(0,\mathcal{T})$.
Then, the time average VoIU (normalized by the duration of time interval) is given by 
\begin{equation}
V_{\mathcal{T}} = \frac{1}{\mathcal{T}} \sum_{i=1}^{N(\mathcal{T})} V_i.
\label{eq:av_value}
\end{equation} 
Without loss of generality we assume that the first packet generation was at the time instant $t_0$ and the observation begins at $t = 0$ with an empty queue and the value $C(0) = C_0$.
Moreover, the observation interval ends with the service completion of $N(\mathcal{T})$ samples, with $N(\mathcal{T}) = max \{n \,|\, t_n \leq \mathcal{T}\}$ denoting the number of arrivals by time $\mathcal{T}$.

The time average value in \eqref{eq:av_value} is an important metric taken into consideration when evaluating the performance of a network of status updates and should be calculated for each case of $f_s(t)$ separately. 
The time average VoIU for the three considered cases can be rewritten as 
\begin{equation}
V_\mathcal{T} =  \frac{N(\mathcal{T})}{\mathcal{T}} \frac{1}{N(\mathcal{T})} \sum^{N(\mathcal{T})}_{i=1} V_i.
\label{eq:V_i2}
\end{equation}
Additionally, defining the steady-state time average arrival rate as 
\begin{equation}
\lambda = \lim_{\mathcal{T} \to \infty} \frac{N(\mathcal{T})}{\mathcal{T}} 
\label{eq:lambda}
\end{equation}
and noticing that $N(\mathcal{T}) \to \infty$ as $\mathcal{T} \to \infty$, and that the sample average will converge to its corresponding stochastic average due to the assumed ergodicity of $V_i$, we conclude with the 
expression 
 \begin{equation}
V = \lim_{\mathcal{T} \to \infty} V_\mathcal{T} = \lambda \:\mathbb{E}[V],
\label{eq:av_V}
\end{equation}
where $\mathbb{E}[\cdot]$ is the expectation operator.
Next, we focus on CoUD and analyze it for each case of $f_s(t)$ separately. 

\section{Cost of Update Delay Analysis}
\label{sec:avCoUD}
We first derive useful results for the general case without
considering specific queueing models.
The time average CoUD of (\ref{eq:cost_t}) in this scenario can be calculated as the  sum of the disjoint $Q_1$, $Q_i$ for $i \geq 2$, and the area of width $T_n$ over the time interval  $(t_n, t'_n)$, denoted by $\tilde{Q}$.
This decomposition yields 
\begin{equation}
C_\mathcal{T} = \frac{Q_1 + \tilde{Q} + \sum^{N(\mathcal{T})}_{i=2}Q_i}{\mathcal{T}}.
\label{eq:Delta_t2}
\end{equation}  
Below we derive the average CoUD for the three cases of the $f_s(t)$ function that we have considered and find the optimum server policy for each one of them.

For $f_s(t)=\alpha t$, the area $Q_i$ for $i \geq 2$ is a trapezoid equal to the difference of two triangles, hence  
\begin{equation}
Q_{P,i} = \frac{1}{2}\alpha(T_i + Y_i)^2 - \frac{1}{2} \alpha T_i^2 = \alpha \big[Y_i T_i + \frac{Y_i^2}{2} \big].
\label{eq:Qi_at}
\end{equation}
%For $\alpha = 1$ CoUD corresponds to the timeliness of each status update arriving and is the so called \emph{age of information}.
Next, for $f_s(t)=e^{\alpha  t} -1$, the area  $Q_i$ yields
\begin{align}
Q_{E,i} =  \int^{t'_i}_{t_{i-1}}  (e^{\alpha (t-t_{i-1})} -1) \,\mathrm{d}t -  \int_{t_i}^{t'_i}  (e^{\alpha  (t - t_i)} -1) \,\mathrm{d}t = \frac{1}{\alpha}  \big[ e^{\alpha (Y_i + T_i)} - e^{\alpha T_i} \big] - Y_i.
\label{eq:Qi_exp(t)}
\end{align}
%\begin{align}
%Q_{E,i} &=  \int^{t'_i}_{t_{i-1}}  (e^{\alpha (t-t_{i-1})} -1) \,\mathrm{d}t -  \int_{t_i}^{t'_i}  (e^{\alpha  (t - t_i)} -1) \,\mathrm{d}t = \notag \\
%&= \frac{1}{\alpha}  \big[ e^{\alpha (Y_i + T_i)} - e^{\alpha T_i} \big] - Y_i.
%\label{eq:Qi_exp(t)}
%\end{align}
And lastly, for $f_s(t)=\log(\alpha t +1)$ we obtain
\begin{equation*}
Q_{L,i} =  \int^{t'_i}_{t_{i-1}}  \log((\alpha (t-t_{i-1})+1) \,\mathrm{d}t -  \int_{t_i}^{t'_i}  \log(\alpha (t-t_i) +1) \,\mathrm{d}t =
\label{eq:Qi_log}
\end{equation*}
\begin{equation}
 =\frac{1}{\alpha}  \big[ (\alpha (Y_i + T_i) +1) \log(\alpha (Y_i + T_i) +1) - (\alpha T_i+1) \log(\alpha T_i +1) \big] - Y_i. 
\label{eq:Qi_log2}
\end{equation}
%\begin{equation}
%\resizebox{.99\hsize}{!}{$ =\frac{1}{\alpha}  \big[ (\alpha (Y_i + T_i) +1) \log(\alpha (Y_i + T_i) +1) - (\alpha T_i+1) \log(\alpha T_i +1) \big] - Y_i. $}
%\label{eq:Qi_log2}
%\end{equation}
The time average CoUD for the three cases can be rewritten as 
\begin{equation}
C_\mathcal{T} = \frac{\bar{Q}}{\mathcal{T}} + \frac{N(\mathcal{T})-1}{\mathcal{T}} \frac{1}{N(\mathcal{T})-1} \sum^{N(\mathcal{T})}_{i=2} Q_i
\label{eq:Delta_3}
\end{equation}
where, $\bar{Q} = Q_1 + \tilde{Q}$ and $\bar{Q}/\mathcal{T}$ is a term that will vanish as $\mathcal{T} \to \infty$. 
Then, similarly to the VoIU analysis, we conclude with the expression 
 \begin{equation}
C = \lim_{\mathcal{T} \to \infty} C_\mathcal{T} = \lambda \:\mathbb{E}[Q].
\label{eq:av_Delta}
\end{equation}
In addition, an alternative formula of the time average CoUD can be written in terms of the stationary distribution $C(t)$ of the CoUD as
 \begin{equation}
C = \lim_{\mathcal{T} \to \infty} C_\mathcal{T} = \int_0^{\infty} f_s(t) \: \mathrm{d}C(t).
\label{eq:av_Delta_v2}
\end{equation}

\section{CoUD and PCoUD computation for the M/M/1 System}
For an M/M/1 system, status updates are generated according to a Poisson process with mean $\lambda$, thus the interarrival times $Y_i$ are independent and identically distributed (i.i.d.) exponential random variables with $\mathbb{E}[Y] = 1/\lambda$.
The service times are i.i.d. exponentially distributed with mean $\mathbb{E}[S]=1/\mu$ and the server utilization is $\rho = \frac{\lambda}{\mu}$.
Furthermore, the distribution of the system time $T$ for the M/M/1 system is given by $\mathbb{P}_T(t) = \mu(1-\rho) e^{-\mu(1-\rho)t}, \:\:\: t \geq 0$, and it represents an exponential probability density function (pdf) with mean $\mathbb{E}[T]=1/(\mu-\rho)$.

\begin{theorem}\label{theorem1}
	For the $f_s(t)=\alpha t$ case, the average CoUD for the M/M/1 system with an FCFS queue discipline is given by
	 \begin{equation}
	C_{P} = \alpha \frac{1}{\mu} \left( 1+\frac{1}{\rho}+\frac{\rho^2}{1-\rho} \right),
	\label{eq:av_C_lin}
	\end{equation}
    and the average PCoUD is given by
   \begin{equation}
   	A_{P} = \alpha \left(\frac{1}{\lambda}+ \frac{1}{\mu-\lambda} \right).
   	\label{eq:av_A_lin}
   \end{equation}
	\begin{proof} The proof is given in Appendix~\ref{Appendix_A'}. \end{proof}	
\end{theorem}
\begin{remark}
For $\alpha=1$ the results in \eqref{eq:av_C_lin} and \eqref{eq:av_A_lin} are the AoI of an M/M/1 queue that is found in \cite{Kaul12_INFOCOM} and the PAoI of an M/M/1 queue that is found in \cite{Modiano15_ISIT}, respectively.
\end{remark}

\begin{corollary}\label{corollary1}
The covariance $Cov[\cdot]$ of the waiting time $W$ and the interarrival time $Y$ is equal to the covariance of the system time $T$ and the interarrival time $Y$, and is given by the negative term 
\begin{align}
Cov[W, Y] &= \mathbb{E}[(W - \mathbb{E}[W])  (Y - \mathbb{E}[Y])] = Cov[T, Y] =\mathbb{E}[(T - \mathbb{E}[T])  (Y - \mathbb{E}[Y])] = \notag \\
& = - \frac{1}{\mu^2}.
\label{eq:covWY}
\end{align}
\end{corollary}

\begin{corollary}\label{corollary2}
	For the $f_s(t)=\alpha  t$ case, the average CoUD for the M/M/1 system with an FCFS queue discipline is upper bounded by 
	\begin{equation}
	C_{P} \leq \lambda \alpha \left( \frac{1}{\lambda} \frac{1}{\mu (1-\rho)} + \frac{1}{\lambda^2} \right).
	\label{eq:av_C_bound}
	\end{equation}
	Let $\bar{Y}$ be a random variable that is i.i.d. with $Y$ and independent of $T$.
	The upper bound for the average CoUD in \eqref{eq:av_C_bound} equals the average PCoUD $\mathbb{E}\left[\alpha (\bar{Y}+T)\right]$.
	\begin{proof} The proof is straightforward from the analysis in Theorem~\ref{theorem1} and is thus omitted. \end{proof}	
\end{corollary}

Next, for the exponential case we have the following theorem.
\begin{theorem}\label{theorem2}
	For the $f_s(t)=e^{\alpha  t} -1$ case, the average CoUD for the M/M/1 system with an FCFS queue discipline is given by
	 \begin{equation}
	C_{E} = \mu (\rho-1) \left(\frac{\alpha (\alpha-(\lambda+\mu))}{(\lambda-\alpha) (\alpha-\mu)^2}+\frac{1}{\alpha-\mu (1-\rho)}+\frac{1}{\mu(1- \rho)}\right),
	\label{eq:av_C_exp}
	\end{equation}
	and the average PCoUD is given by
	\begin{equation}
	A_{E} = \frac{\alpha \left(3 a^2 \mu-a^3+\alpha \left(\lambda^2-\lambda \mu-3 \mu^2\right)+\mu^3\right)}{(\lambda-\alpha) (\alpha-\mu)^2 (\mu-\lambda-\alpha)},
	\label{eq:av_A_exp}
	\end{equation}
	where $\alpha<\lambda$ and $\alpha<\mu-\lambda$.
	\begin{proof} The proof is given in Appendix~\ref{Appendix_B'}. \end{proof}	
\end{theorem}

\begin{corollary}\label{corollary3}
	For the $f_s(t)=e^{\alpha  t} -1$ case, the average CoUD for the M/M/1 system with an FCFS queue discipline is upper bounded by
	\begin{equation}
	C_{E} \leq \lambda \frac{1}{\alpha} \left( \frac{\mu(1-\rho)}{\alpha-\mu(1-\rho)} \left( \frac{\lambda}{\alpha-\lambda}+1 \right) \right) -1,
	\label{eq:av_A_exp_bound}
	\end{equation}
	where $\alpha<\lambda$ and $\alpha<\mu-\lambda$.
	Let $\bar{Y}$ be a random variable that is i.i.d. with $Y$ and independent of $T$. The upper bound for the average CoUD in \eqref{eq:av_A_exp_bound} equals the upper bound for the average PCoUD $\mathbb{E}\left[e^{\alpha (\bar{Y}+T)} -1\right]$.
	\begin{proof} The proof is given in Appendix~\ref{Appendix_C'}. \end{proof}	
\end{corollary}

Finally, for the logarithmic case we have the following theorem.
\begin{theorem}\label{theorem3}
	For the $f_s(t)=\log{(\alpha  t+1)}$ case, the average CoUD for the M/M/1 system with an FCFS queue discipline is given by
	\begin{align}
	C_{L} &=  
	\frac{1}{\alpha (\lambda-\mu)^2} \Bigg(
	e^{-\frac{\mu \rho}{\alpha}} \Big( \mu (1-\rho) Ei\left[-\frac{\mu}{a}\right] \left(\alpha \mu+\lambda^2-\lambda \mu\right) e^{\frac{\mu (\rho+1)}{\alpha}}-\alpha \mu^2 (1-\rho)  Ei\left[-\frac{\lambda}{\alpha}\right] e^{\frac{\lambda+\mu \rho}{\alpha}} \notag \\
	&-\alpha e^{\mu/\alpha} (\lambda-\mu)^2 Ei\left[-\frac{\mu (1-\rho)}{\alpha}\right] \Big) 
    -\alpha \lambda (1-\rho) (\mu-\lambda) \Bigg),
	\label{eq:av_C_log}
	\end{align}
	and the average PCoUD is given by
	\begin{align}
	A_{L} &= \frac{1}{\alpha \lambda (\lambda-\mu)} \Bigg( e^{-\frac{\mu \rho}{\alpha}} \Big( \alpha (\mu-\lambda) \left(\mu e^{\mu/\alpha} Ei \left[-\frac{\mu (1-\rho)}{\alpha}\right] + \lambda e^{\frac{\mu \rho}{\alpha}}\right) \notag \\
	&- Ei \left[-\frac{\mu}{\alpha}\right] \left(\alpha \left(\lambda^2-\lambda \mu+\mu^2\right)+\lambda \mu (\lambda-\mu)\right) e^{\frac{\mu (\rho+1)}{\alpha}} \Big)+\alpha \lambda \mu e^{\lambda/a} Ei \left[-\frac{\lambda}{\alpha}\right] \Bigg),
	\label{eq:av_A_log}
	\end{align}
	where $Ei$ denotes the exponential integral defined as 
	\begin{equation}
	Ei[x] = -\int_{-x}^{\infty} \frac{e^{-t}}{t} \: \mathrm{d}t.
	\label{eq:exp_integr}
	\end{equation}
	\begin{proof} The proof is along the lines of the proof in Appendix~\ref{Appendix_B'} and is therefore omitted. \end{proof}	
\end{theorem}

\begin{corollary}\label{corollary4}
	For the $f_s(t)=\log{(\alpha  t+1)}$ case, the average CoUD for the M/M/1 system with an FCFS queue discipline is upper bounded by
	\begin{equation}
	C_{L} \leq \lambda \frac{e^{\frac{\mu-\lambda}{\alpha}} \left( \lambda Ei \left[ \frac{\lambda-\mu}{\alpha}\right] +(\lambda-\mu) Ei \left[-\frac{\lambda}{\alpha}\right] e^{\frac{2 \lambda-\mu}{\alpha}}\right)}{\lambda (\mu-2 \lambda)},
	\label{eq:av_C_log_bound}
	\end{equation}
	where $Ei$ denotes the exponential integral defined in \eqref{eq:exp_integr}.
    The limit of \eqref{eq:av_C_log_bound} as $\lambda$ approaches $\mu/2$ gives 
    \begin{equation}
    C_{L} \leq \lambda \frac{2 \alpha-e^{\frac{\mu}{2 \alpha}} (2 \alpha-\mu) Ei \left(-\frac{\mu}{2 \alpha}\right)}{\alpha \mu}.
    \end{equation} 
	Let $\bar{Y}$ be a random variable that is i.i.d. with $Y$ and independent of $T$.
	The average CoUD upper bound in \eqref{eq:av_C_log_bound} equals the lower bound for the average PCoUD
	$\mathbb{E}\left[\log(\alpha (\bar{Y}+T)+1)\right]$.
     \begin{proof} The proof is given in Appendix~\ref{Appendix_D'}. \end{proof}	
\end{corollary}

\section{Value of Information of Update Computation for the M/M/1 System}
\label{sec:avVoI}
Following the same procedure as in the CoUD metric, we compute the average VoIU given by (\ref{eq:av_V}), for the M/M/1 system with an FCFS queue discipline.

\begin{theorem}\label{theorem4}
	For the $f_s(t)=\alpha t$ case, the average VoIU for the M/M/1 system with an FCFS queue discipline is approximated by
    \begin{equation}
    V_P = \lambda \frac{(1-\rho)}{2\rho} \: {}_2 F_1 \left(1,2{;}3{;}2-\frac{1 }{\rho}\right),
    \label{eq:av_V_lin}
    \end{equation}
    where ${}_2 F_1$ is the hypergeometric function defined by the power series
    \begin{equation}
    {}_2 F_1 \left(a,b{;}c{;}z\right) = \sum_{n=0}^{\infty} \frac{(a)_n (b)_n}{(c)_n} \frac{z^n}{n!},
    \end{equation}
    for $|z| < 1$ and by analytic continuation elsewhere.
    Here $(q)_n$ is the Pochhammer symbol, which is defined by 
    \begin{align}
    (q)_n  =\begin{cases} 
    1&\mbox{, if } n = 0 \\
    q (q+1) \cdots (q+n-1)&\mbox{, if } n >0.
    \end{cases} 
    \end{align}
   \begin{proof} The proof is given in Appendix~\ref{Appendix_E'}. \end{proof}	
\end{theorem}

\begin{remark}
	For the cases $f_s(t)=e^{\alpha  t} -1$ and $f_s(t)=\log(\alpha t+1)$, we compute numerically the expected values $\mathbb{E}\big[V_E\big]$ and $\mathbb{E}\big[V_L\big]$, respectively, by	
	\begin{align}
	& \mathbb{E}\left[\frac{e^{\alpha (Y+T)}-e^{\alpha T}}{e^{\alpha (Y+T)}-1}\right] =  \int_0^\infty \mathbb{E}\left[\frac{e^{\alpha (Y+T)}-e^{\alpha T}}{e^{\alpha (Y+T)}-1}\Big| Y=y\right] \lambda e^{-\lambda y} \:\mathrm{d}y,
	\label{eq:av_V_exp}
	\end{align}
	%\begin{align}
	%& \mathbb{E}\left[\frac{e^{\alpha (Y+T)}-e^{\alpha T}}{e^{\alpha (Y+T)}-1}\right] = \notag  \\
	%& \quad\quad = \int_0^\infty \mathbb{E}\left[\frac{e^{\alpha (Y+T)}-e^{\alpha T}}{e^{\alpha (Y+T)}-1}\Big| Y=y\right] \lambda e^{-\lambda y} \:\mathrm{d}y,
	%\label{eq:av_V_exp}
	%\end{align}
	\begin{align}
	& \mathbb{E}\left[\frac{\log(\alpha (Y + T) +1) - \log(\alpha T+1)}{\log(\alpha (Y + T) +1)}\right] = \notag  \\
	& = \int_0^\infty \mathbb{E}\left[\frac{\log(\alpha (Y + T) +1) - \log(\alpha T+1)}{\log(\alpha (Y + T) +1)}\Big| Y=y\right] \lambda e^{-\lambda y} \:\mathrm{d}y.
	\label{eq:av_V_log}
	\end{align}
	%\begin{align}
	%& \mathbb{E}\left[\frac{\log(\alpha (Y + T) +1) - \log(\alpha T+1)}{\log(\alpha (Y + T) +1)}\right] = \notag  \\
	%& = \int_0^\infty \mathbb{E}\left[\frac{\log(\alpha (Y + T) +1) - \log(\alpha T+1)}{\log(\alpha (Y + T) +1)}\Big| Y=y\right] \notag  \\
	%&\times \lambda e^{-\lambda y} \:\mathrm{d}y.
	%\label{eq:av_V_log}
	%\end{align}
	
	Note that the correlation between $Y$ and $T$ is neglected however besides the analytical results we provide a simulation evaluation in Section~\ref{sec:results}.
\end{remark}

\begin{figure}[t!]\centering
	\centering
	\includegraphics[draft=false,scale=.5]{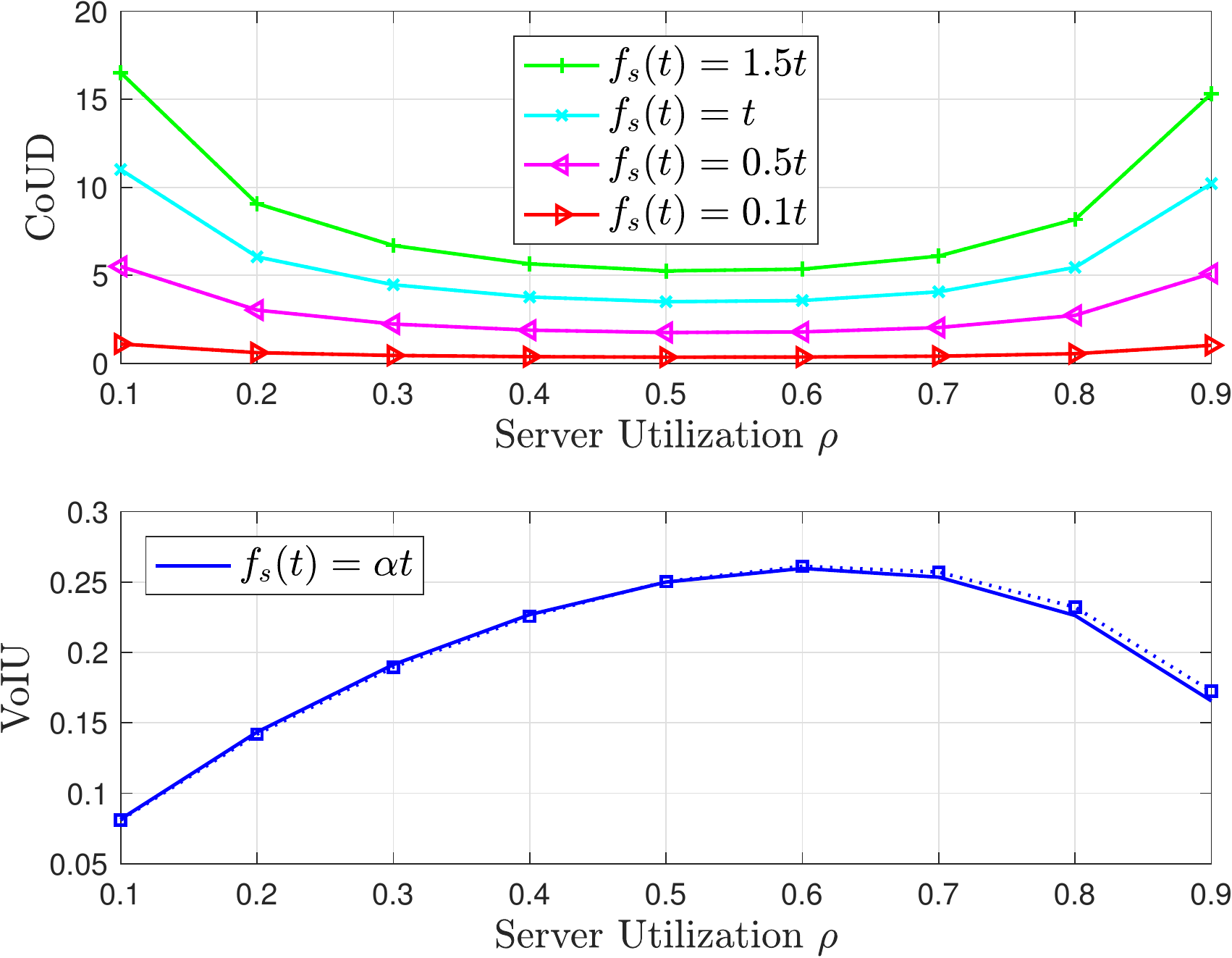}
	\vspace{-3mm}
	\caption{Average CoUD and VoIU vs. the server utilization for the M/M/1 system with $\mu=1$, linear case.}
	\label{fig:CoUDandVoI_vs_utilization_linear}
\end{figure}

\begin{figure}[t!]\centering
	\centering
	\includegraphics[draft=false,scale=.5]{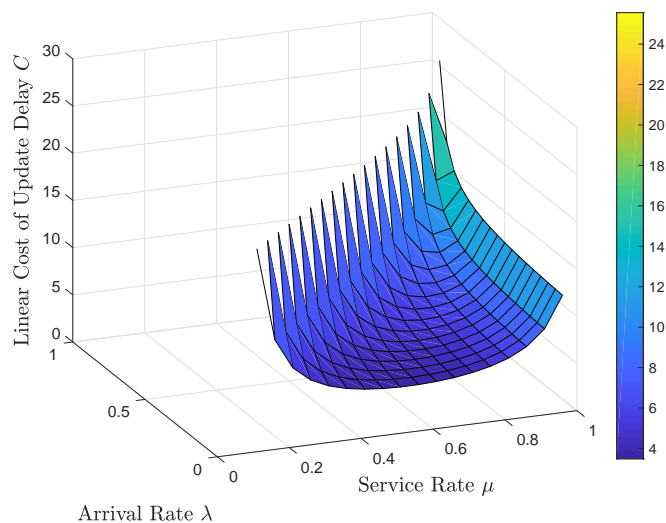}
	\vspace{-3mm}
	\caption{Average CoUD vs. the service rate vs. the arrival rate for the M/M/1 system, linear case with $\alpha=1$.}
	\label{fig:CoUD_vs_mu_vs_lambda_linear}
\end{figure}

\begin{figure}[t!]\centering
	\centering
	\includegraphics[draft=false,scale=.5]{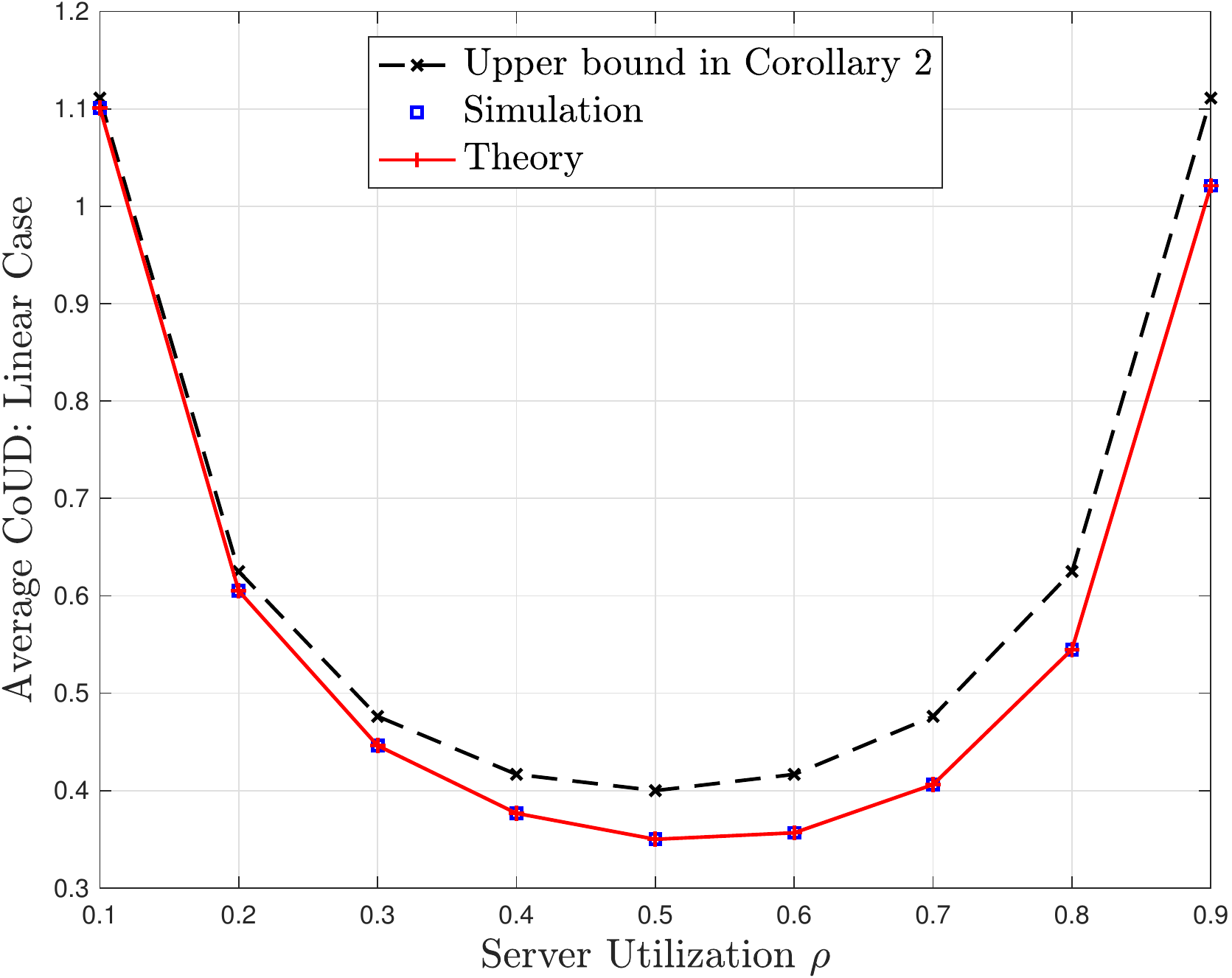}
	\vspace{-3mm}
	\caption{Average CoUD vs. the server utilization for the M/M/1 system with $\mu=1$, linear case with $\alpha=0.1$.}
	\label{fig:CoUD_vs_utilization_linear}
\end{figure}

\section{Numerical Results}
\label{sec:results}
In this section, we evaluate the performance of the system in terms of the CoUD, the PCoUD, and the VoIU metrics, as calculated in the previous section.
In addition, we develop a MATLAB-based event driven 
simulator where each case runs for $10^6$ timeslots, to validate the analytical results.
We consider an M/M/1 system model with average arrival rate $\lambda$, average service rate $\mu$, and server utilization $\rho=\frac{\lambda}{\mu}$.

\vspace{-3mm}
In Fig.~\ref{fig:CoUDandVoI_vs_utilization_linear}, we illustrate the variation of the average CoUD and VoIU with the server utilization $\rho$, for the linear case.
Solid and dotted lines with markers correspond to the analytical and simulated results, respectively.
Recall that for $f_s(t)=\alpha  t$ the VoIU is independent of the parameter $\alpha$, therefore for multiple CoUD curves corresponds only one VoIU curve.
This indicates that if the cost per time unit is linearly increased, higher cost leads to higher average CoUD, but the same average VoIU.
This is because we assign to each unit of time the same cost.
Increasing $\alpha$ results in a proportional increase of the average CoUD, however the optimal server policy is the same for every function.
In particular, differentiating \eqref{eq:av_C_lin} with respect to $\rho$ and setting $\partial C_P/\partial \rho=0$, we obtain the optimal utilization $\rho^*= 0.53101$.
Moreover, the optimal policy with respect to VoIU is different than the one for CoUD with the former being greater. 
Differentiating \eqref{eq:av_V_lin} with respect to $\rho$ and then setting $\partial V_P/\partial \rho=0$, we obtain the optimal utilization $\rho^*= 0.614369$.
\emph{This policy utilizes the network resources to the maximum while ensuring that the freshness of the information at the destination remains at a close-to-optimal value.}

In Fig.~\ref{fig:CoUD_vs_mu_vs_lambda_linear}, we plot the average CoUD versus the arrival rate and the service rate of the queue for the linear cost function, with $\alpha=1$.
In the critical points where the stability condition of the queue is violated implying infinite queueing delay, we get the illustrated sawtooth pattern.

\begin{figure*}[t!]
	\centering
	\subfloat[][]{
		%\rule{0.3\linewidth}{3cm}
		%\input{sawtooth_linear.tex}
		\includegraphics[draft=false,scale=.45]{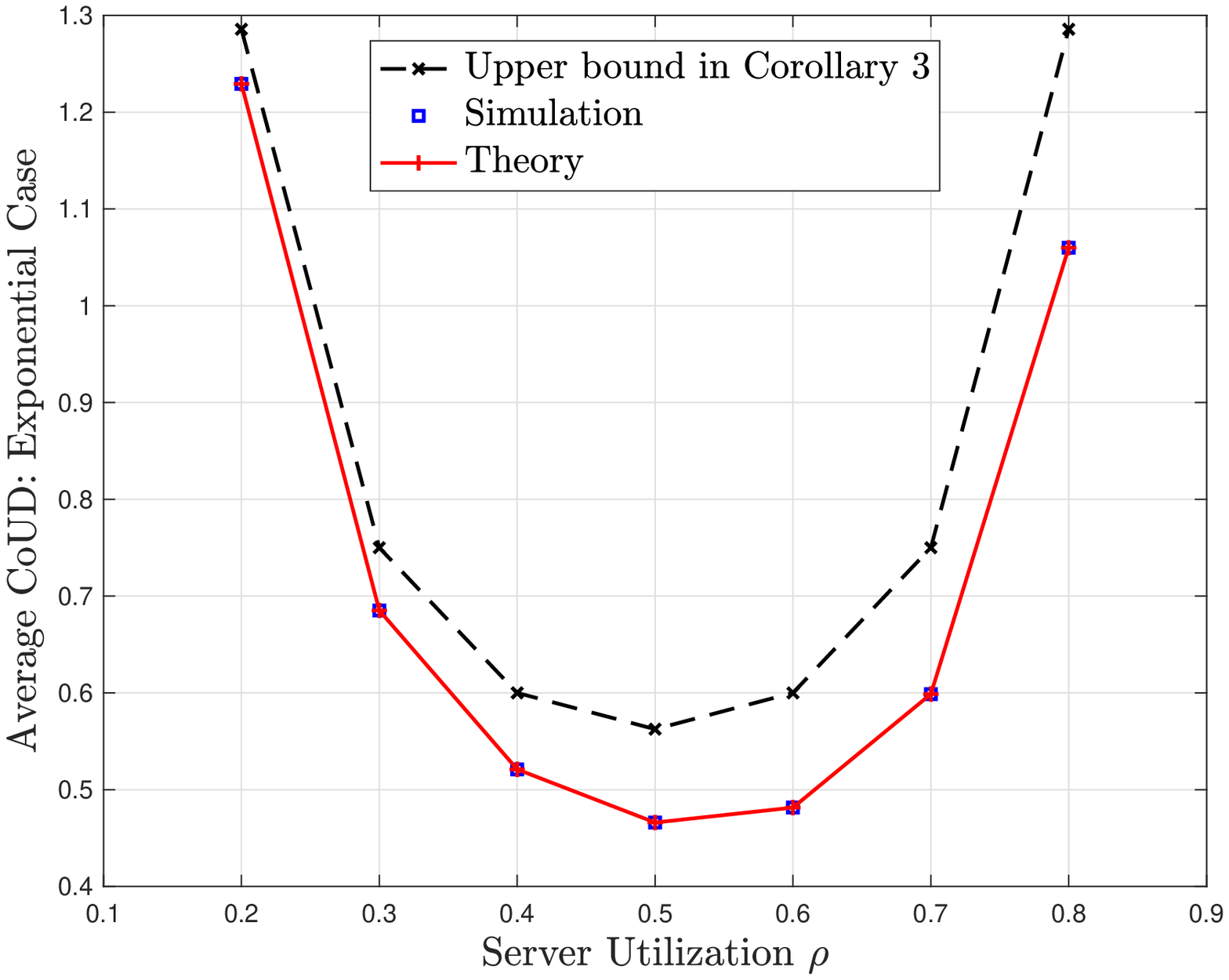}
		\label{fig:CoUD_vs_utilization_exponential}
	}
	\subfloat[][]{
		%\rule{0.3\linewidth}{3cm}
		\includegraphics[draft=false,scale=.45]{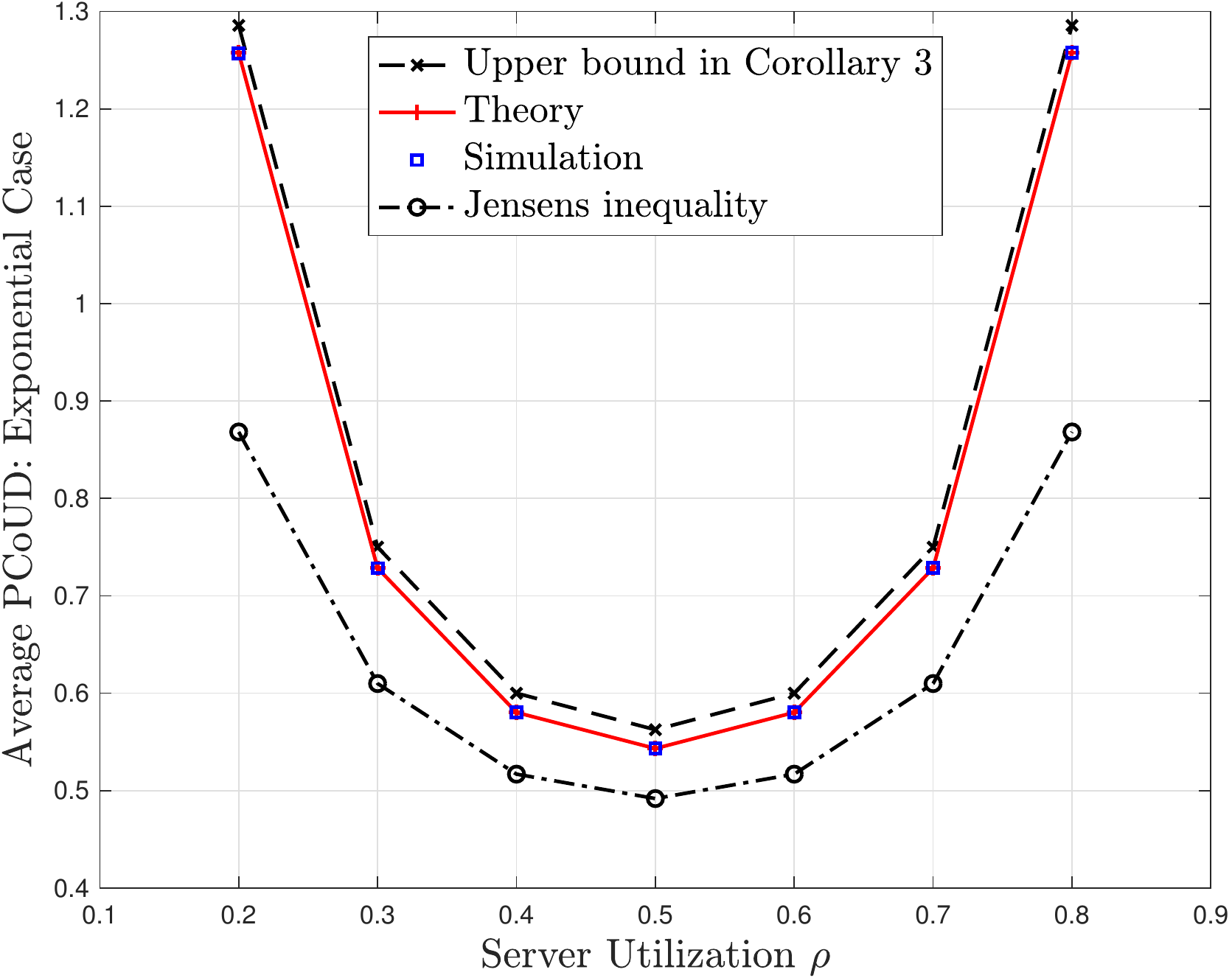}
		\label{fig:PCoUD_vs_utilization_exponential}
	}
   \vspace{-3mm}
	\caption{Average CoUD (left) and PCoUD (right) vs. the server utilization for the M/M/1 system with $\mu=1$, exponential case with $\alpha=0.1$.}
\end{figure*}

\begin{figure*}[t!]
	\centering
	\subfloat[][]{
		%\rule{0.3\linewidth}{3cm}
		%\input{sawtooth_linear.tex}
		\includegraphics[draft=false,scale=.45]{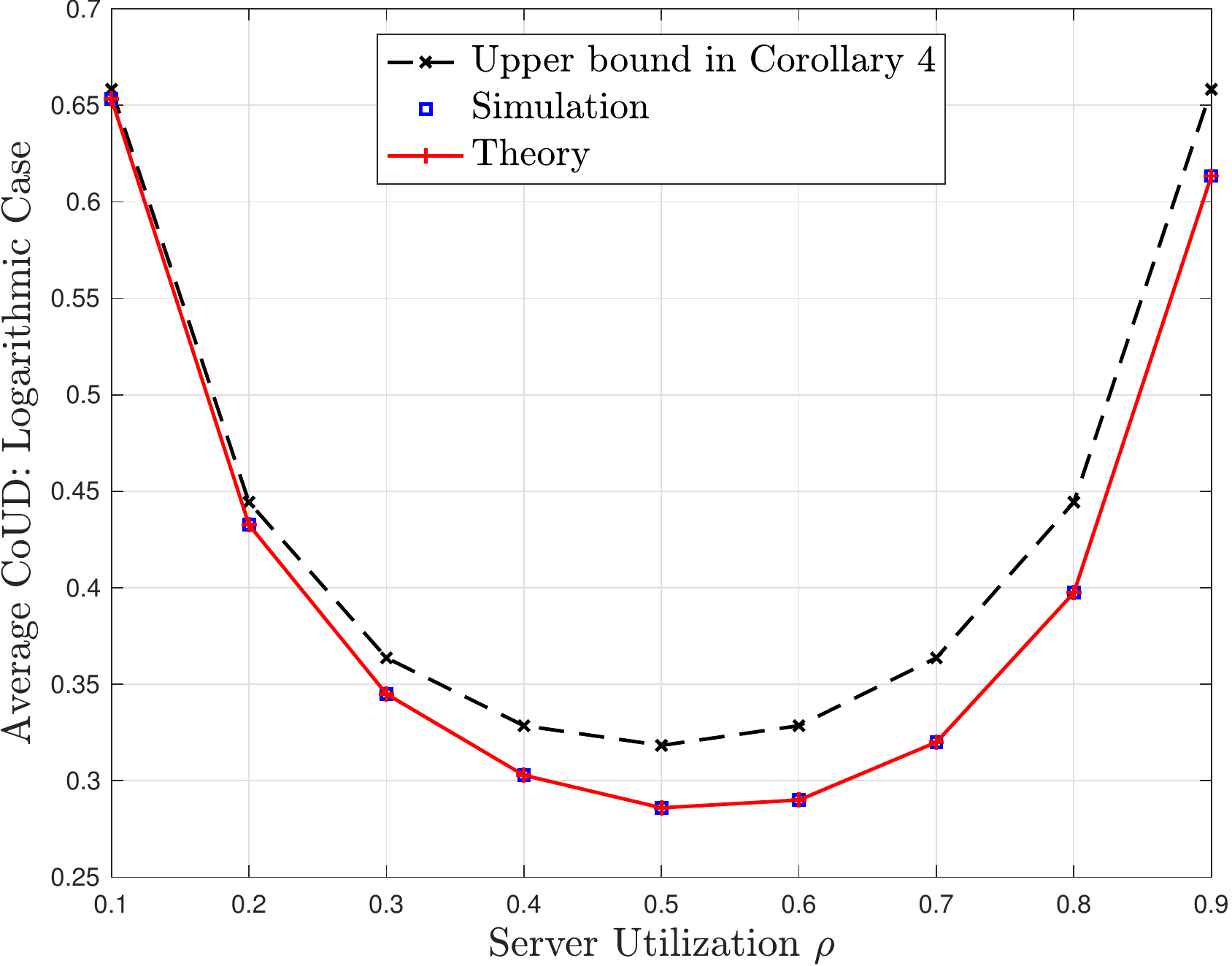}
		\label{fig:CoUD_vs_utilization_logarithmic}
	}
	\subfloat[][]{
		%\rule{0.3\linewidth}{3cm}
		\includegraphics[draft=false,scale=.45]{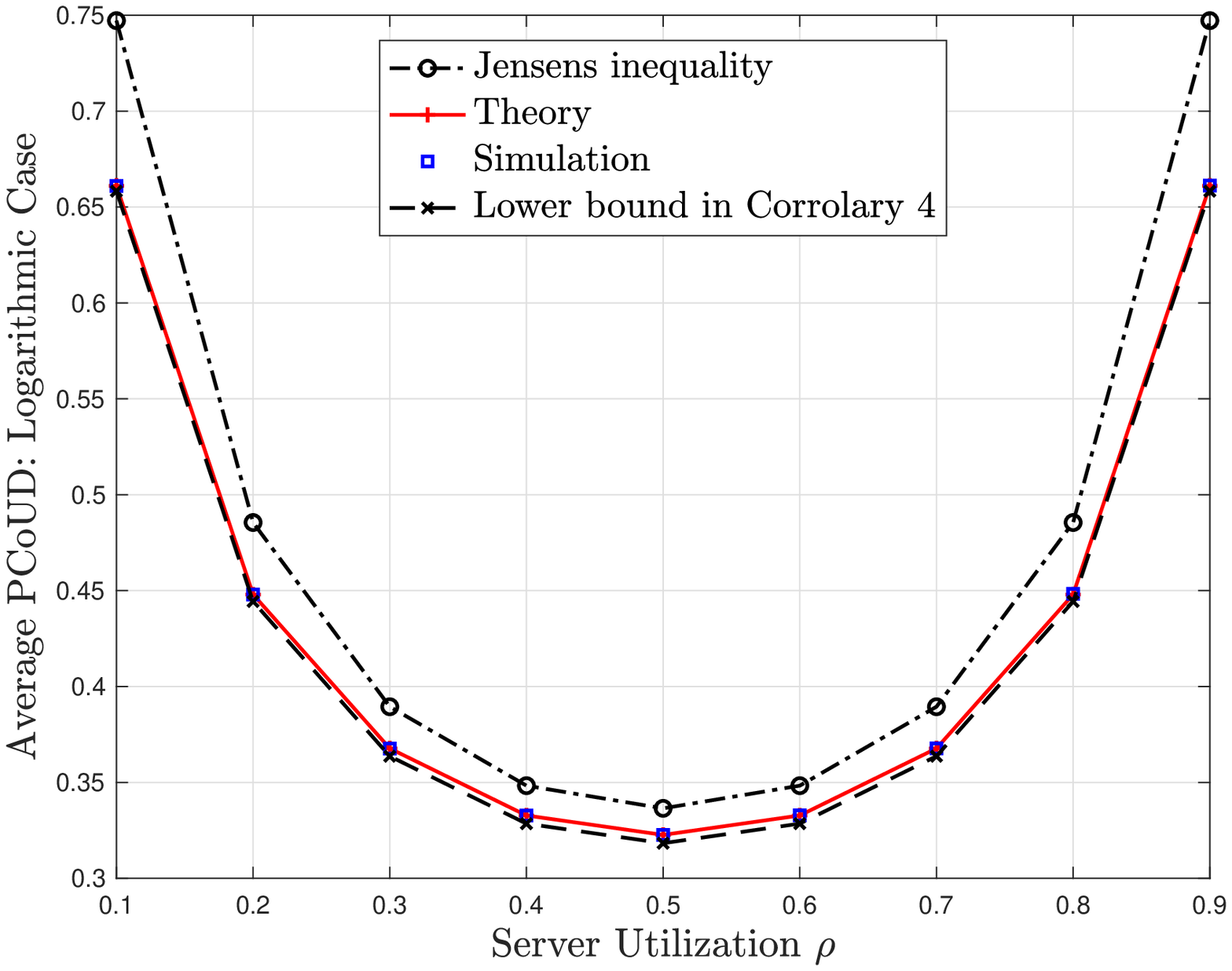}
		\label{fig:PCoUD_vs_utilization_logarithmic}
	}
    \vspace{-3mm}
	\caption{Average CoUD (left) and PCoUD (right) vs. the server utilization for the M/M/1 system with $\mu=1$, logarithmic case for $\alpha=0.1$.}
\end{figure*}

In Fig.~\ref{fig:CoUD_vs_utilization_linear}-\ref{fig:PCoUD_vs_utilization_logarithmic}, we examine the tightness of our bounds and verify the analytical results through simulations. 
The average CoUD and PCoUD are depicted as functions of the server utilization $\rho$, with $\mu=1$, where solid lines, dashed lines, and markers, correspond to the exact results, the bounds in Corollaries 2, 3, and 4, and the simulated CoUD and PCoUD, respectively.
Moreover, we plot upper and lower bounds that derive from Jensen's inequality that states the following:
If $X$ is a random variable and $\phi$ is a convex function, then ${\displaystyle \varphi \left(\mathbb{E} [X]\right)\leq \mathbb{E} \left[\varphi (X)\right]}$.
We observe that our bounds are tight, especially to the PCoUD.
In the case of the non-linear CoUD, increasing $\alpha$ results in a non-proportional increase of the average CoUD, therefore the optimal server policy changes depending on $\alpha$.
For instance, differentiating \eqref{eq:av_C_exp} with respect to $\lambda$ and setting $\partial C_E/\partial \lambda=0$, we obtain the optimal utilization $\rho^*= 0.529098$ for $\alpha=0.1$, and $\rho^*= 0.521131$ for $\alpha=0.3$.
For the logarithmic case, differentiating \eqref{eq:av_C_log} with respect to $\lambda$ and setting $\partial C_L/\partial \lambda=0$, we obtain the optimal utilization $\rho^*= 0.531717$ for $\alpha=0.1$, and $\rho^*= 0.532039$ for $\alpha=0.3$.
However, in the case of the PoUD we find that for all functions PCoUD is minimized by $\rho^*= 0.5$.

\begin{figure*}[t!]
	\centering
	\subfloat[][]{
		%\rule{0.3\linewidth}{3cm}
		%\input{sawtooth_linear.tex}
		\includegraphics[draft=false,scale=.45]{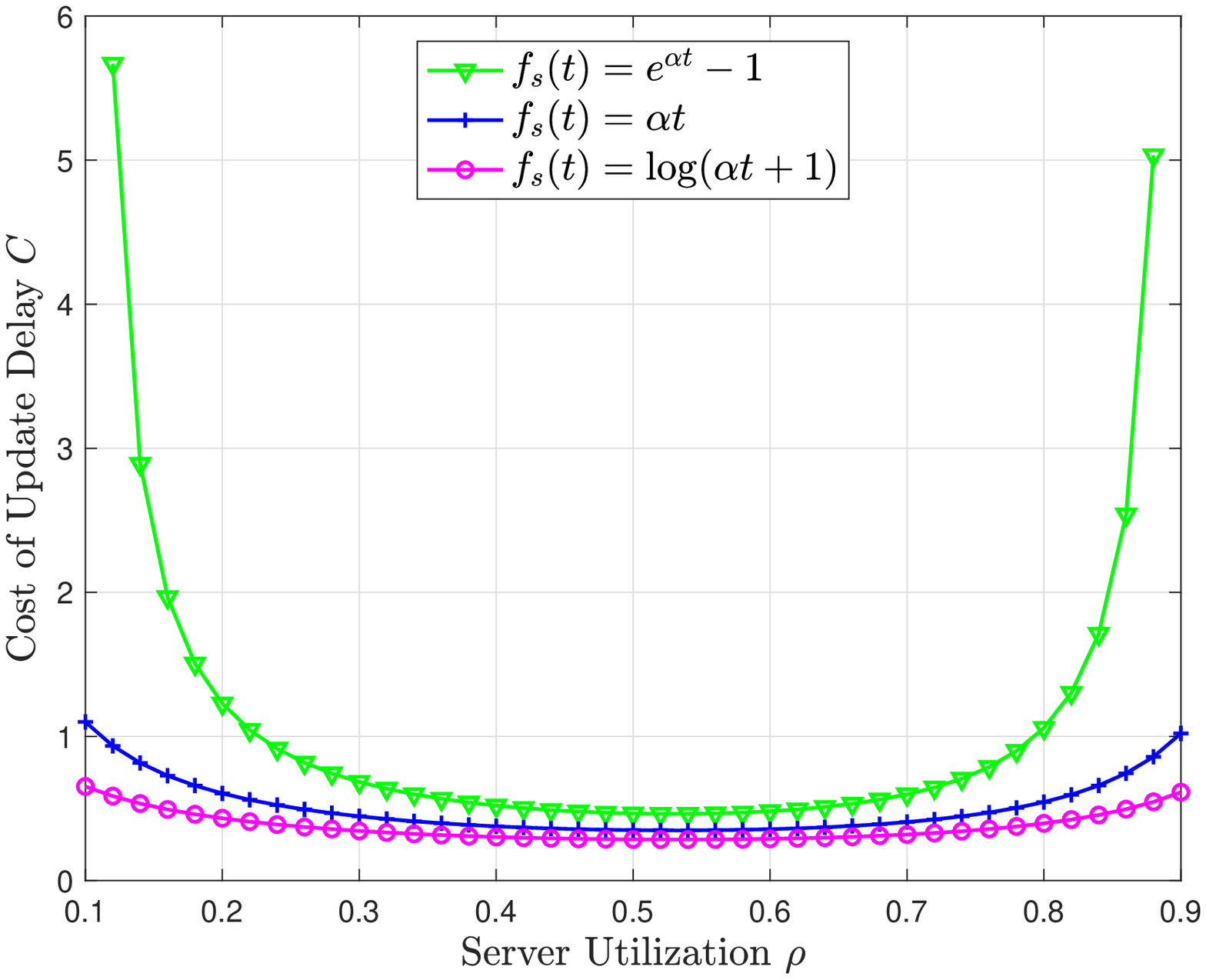}
		\label{fig:CoUD_vs_utilization_casesNEW}
	}
	\subfloat[][]{
		%\rule{0.3\linewidth}{3cm}
		\includegraphics[draft=false,scale=.45]{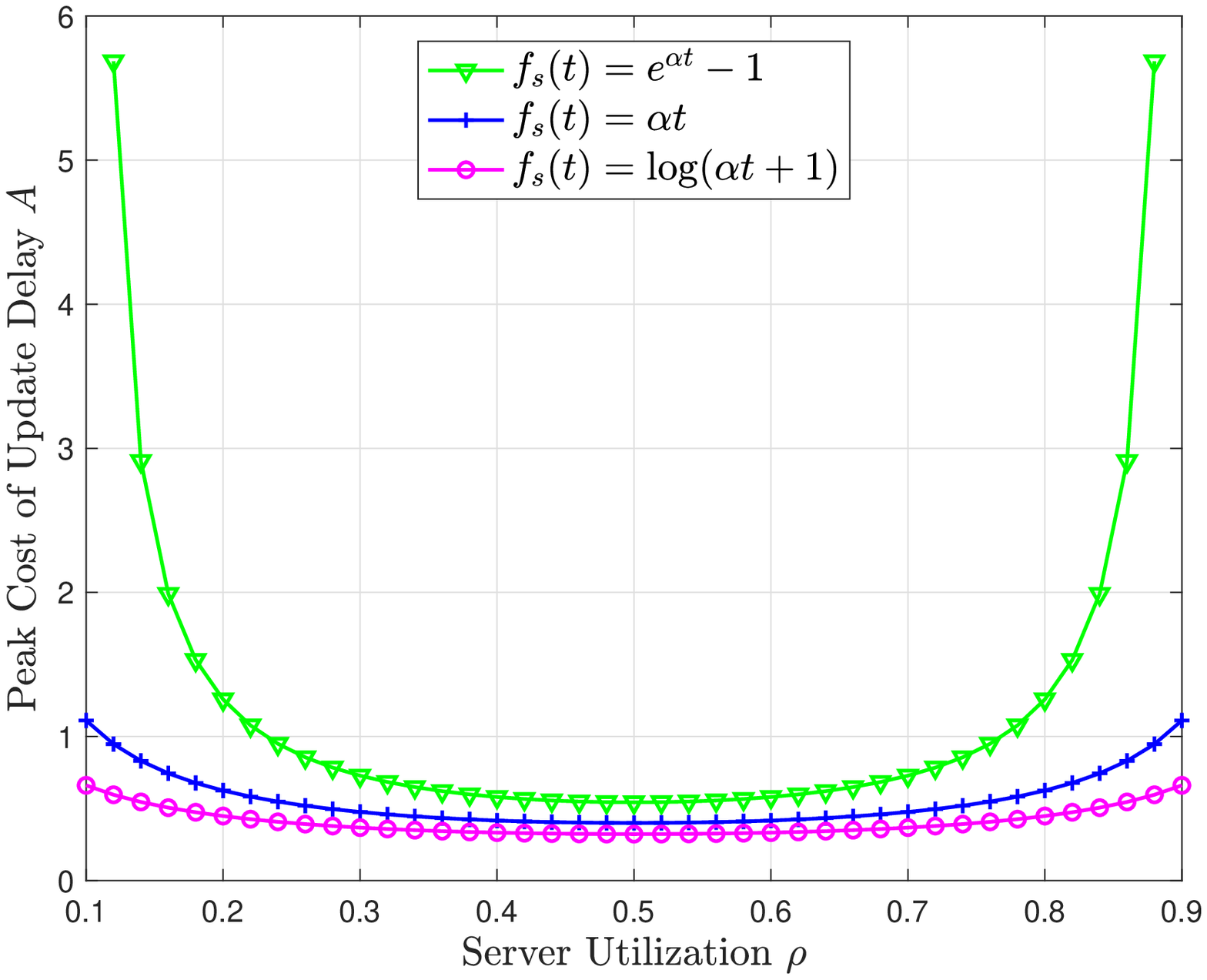}
		\label{fig:PCoUD_vs_utilization_cases}
	}
    \vspace{-3mm}
	\caption{Average CoUD (left) and PCoUD (right) vs. the server utilization for the M/M/1 system with $\mu=1$, and different function cases for $\alpha=0.1$.}
	\label{fig:CoUD_PCoUD_vs_utilization_cases}
\end{figure*}

\begin{figure*}[t!]
	\centering
	\subfloat[][]{
		%\rule{0.3\linewidth}{3cm}
		%\input{sawtooth_linear.tex}
		\includegraphics[draft=false,scale=.45]{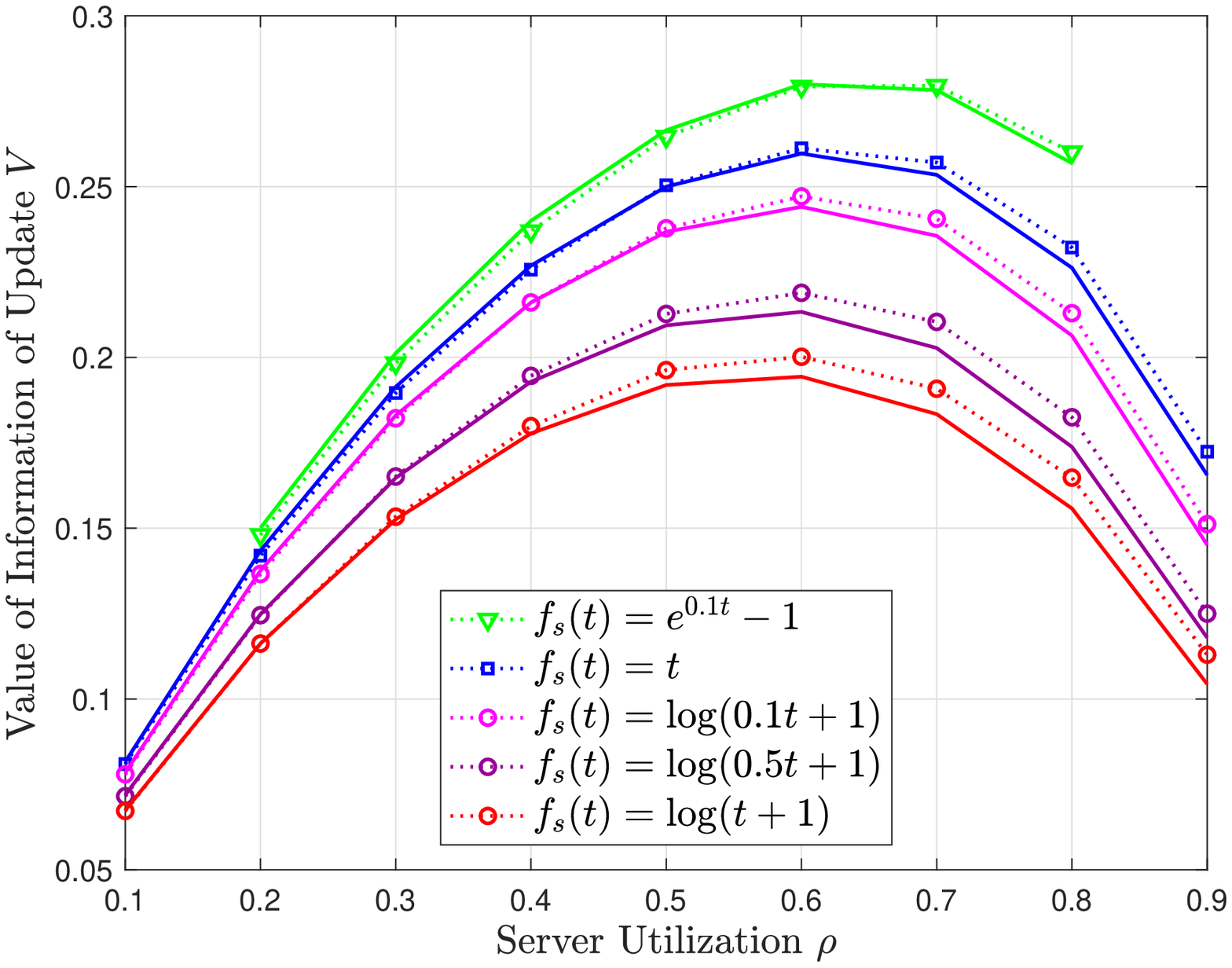}
		\label{fig:VoI_vs_utilization}
	}
	\subfloat[][]{
		%\rule{0.3\linewidth}{3cm}
		\includegraphics[draft=false,scale=.45]{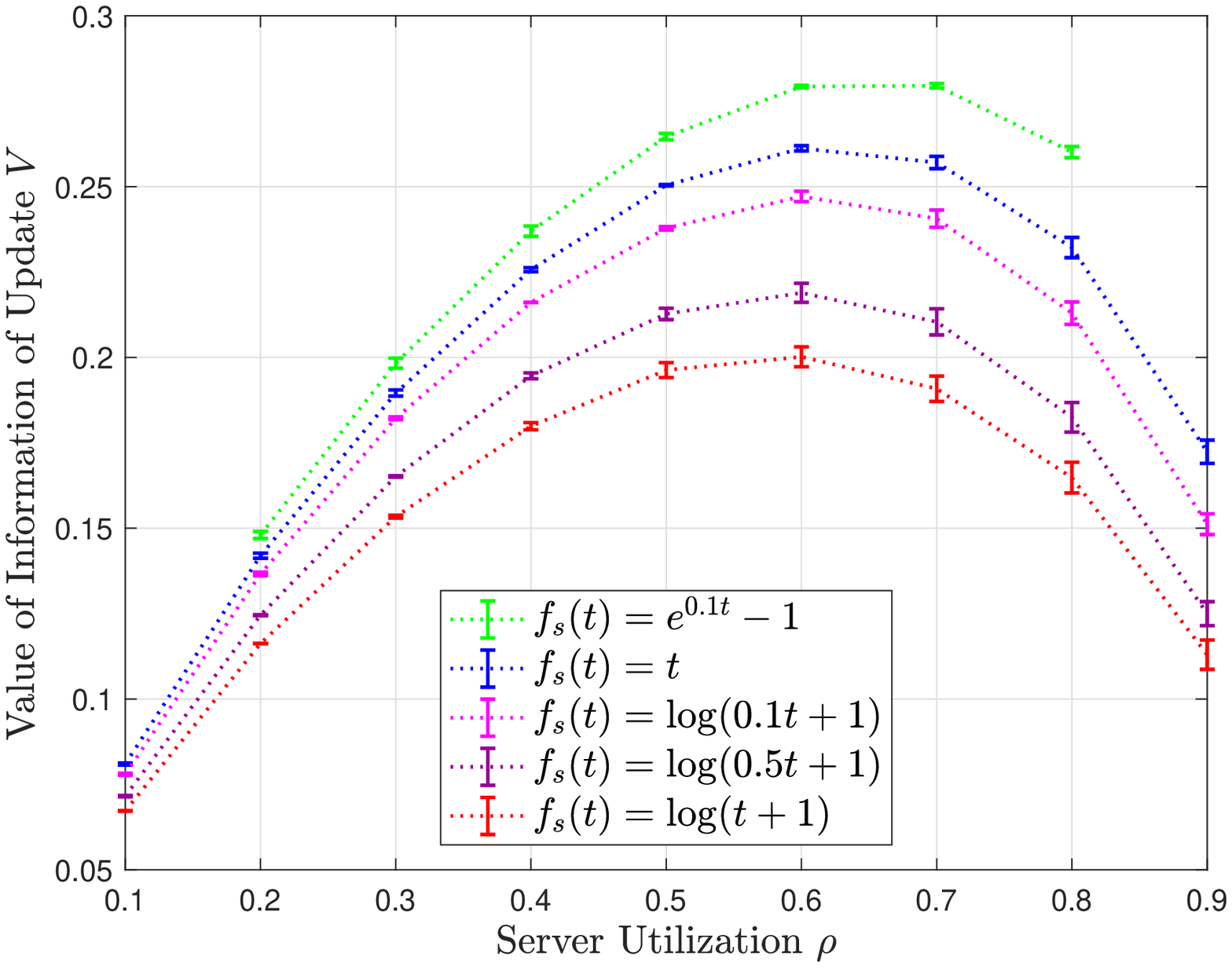}
		\label{fig:VoI_vs_utilization2}
	}
   \vspace{-3mm}
	\caption{Average VoIU vs. the server utilization for the M/M/1 system with $\mu=1$, and different function cases.}
\end{figure*}

In Fig.~\ref{fig:CoUD_PCoUD_vs_utilization_cases}, the average CoUD and PCoUD are depicted as functions of the server utilization for the linear, exponential, and logarithmic functions with parameter $\alpha=0.1$.
Solid lines and dotted lines with markers correspond to the analytical and simulated results, respectively.
All three functions have similar behaviour, with the minimum CoUD achieved when $\rho \approx 0.5$.
Over all values of $\rho$, the exponential $f_s$ yields the highest CoUD, followed by the linear $f_s$ and then the logarithmic $f_s$, that is, $C_E>C_P>C_L$. % with strict inequality.
However, as $\rho$ deviates from the optimum, we see that the exponential function becomes sharper than the linear function, and the logarithmic function becomes smoother.
For smaller utilizations where status updates are not frequent enough and for higher utilization where packets spend more time in the system due to backlogs, CoUD is increased.
%For this increase each function sets its own cost per time unit resulting in more rapid growth for the exponential average CoUD and less intense growth for the logarithmic CoUD.
The difference in this increase is due to the fact that each function sets its own cost per time unit resulting in more rapid growth for the exponential average CoUD and less intense growth for the logarithmic CoUD.

\begin{figure}[t!]\centering
	\centering
	\includegraphics[draft=false,scale=.5]{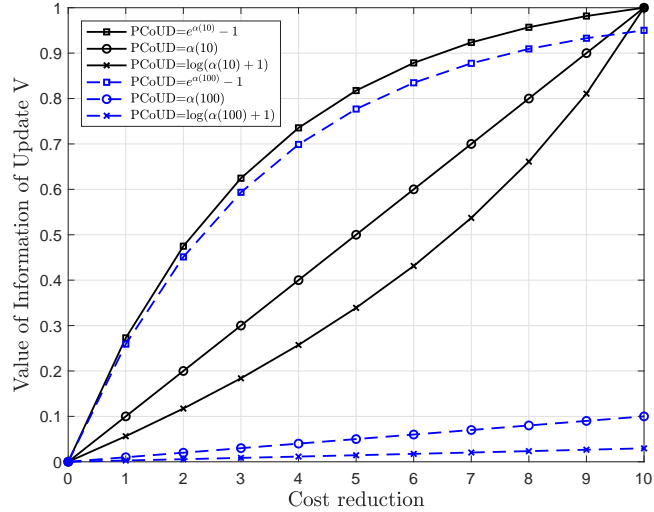}
	\vspace{-3mm}
	\caption{The VoIU vs. the cost reduction, for $\alpha=0.3$.}
	\label{fig:value_vs_drop}
\end{figure}

Fig.~\ref{fig:VoI_vs_utilization} presents the numerical evaluation of the quantities $V_P$, $V_E$, and $V_L$, for three values of the parameter $\alpha$, $0.1$, $0.5$, and $1$.
Solid lines and dotted lines with markers correspond to the analytical and simulated VoIU, respectively.
As we shift from $f_s(t)= \log(t+1)$ to $f_s(t)=e^{0.1t}-1$, VoIU becomes greater over all $\rho$ and all functions follow a similar behaviour.
%, except for $f_s(t)=e^{t}-1$ and $f_s(t)=e^{0.5t}-1$.
%For these functions, due to the fact that CoUD is increasing extremely fast, VoI gets its maximum value for high server utilization $\rho$.
For all cases, the maximum VoIU is achieved when $\rho \approx 0.6$.
Note that VoIU is directly related to CoUD.
However, taking the linear function as a point of reference, the analysis of the average CoUD and VoIU indicates that choosing an exponential function would result in higher CoUD and VoIU, while choosing the logarithmic function would result in lower CoUD and VoIU.
This tradeoff considers two objectives: (i) timeliness, (ii) timeliness and transmission resources (i.e., bandwidth).
In Fig.~\ref{fig:VoI_vs_utilization2} we draw a vertical absolute-difference bar at each server utilization value.
The values of the absolute difference between the analytical and the simulated results determine the length of each bar above and below the server utilization points.

In Fig.~\ref{fig:value_vs_drop} we plot the VoIU vs the cost reduction $D_k$, depicted in Fig.~\ref{fig:linear_value}-\ref{fig:logarithmic_value},
for two values of PCoUD and the three $f_s$ cases, for $\alpha=0.3$.
Specifically, given the event that the sum of the interarrival time and system time of packet $k$ is $Y_k + T_k=10$ or $Y_k + T_k=100$, we are interested in the effect that the cost reduction would have on VoIU.
In the linear case, VoIU increases linearly with the cost reduction both for PCoUD=$f_s(10)$ and PCoUD=$f_s(100)$. 
In case the linear PCoUD and cost reduction are of the same order of magnitude, we observe that VoIU assumes values over its entire range, while in the case where the cost reduction is an order of magnitude lower than the value of PCoUD, the VoIU ranges from 0 to 0.1.
In the exponential and logarithmic cases however, we observe that VoIU increases with the cost reduction inversely proportional to the corresponding function, while maintaining the monotonicity.
%This is an important observation that shows that VoI is indeed a different metric than CoUD with a duality association.
%This duality comes from the fact that 
%This is due to the fact that
Note also that the exponential VoIU for PCoUD=$f_s(100)$ is relatively close to the exponential VoIU for PCoUD=$f_s(10)$, which indicates the necessity of newly received status updates even when 
the decrease in CoUD is small compared to PCoUD. 
For fixed cost reduction, the exponential function has the largest value of VoI and the logarithmic function the smallest, as noted earlier.

\begin{figure}[t!]\centering
	\centering
	\includegraphics[draft=false,scale=.5]{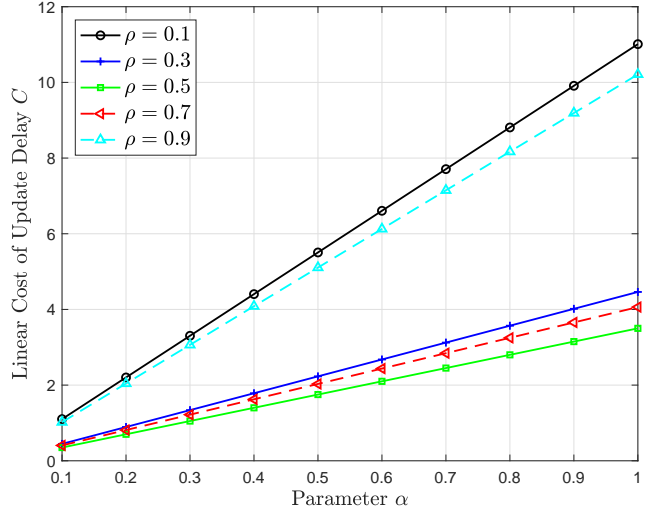}
	\vspace{-3mm}
	\caption{Average CoUD vs. the parameter $\alpha$ for the M/M/1 system with $\mu=1$, linear case.}
	\label{fig:lin_CoUD_vs_alpha}
\end{figure}

\begin{figure}[t!]\centering
	\centering
	\includegraphics[draft=false,scale=.5]{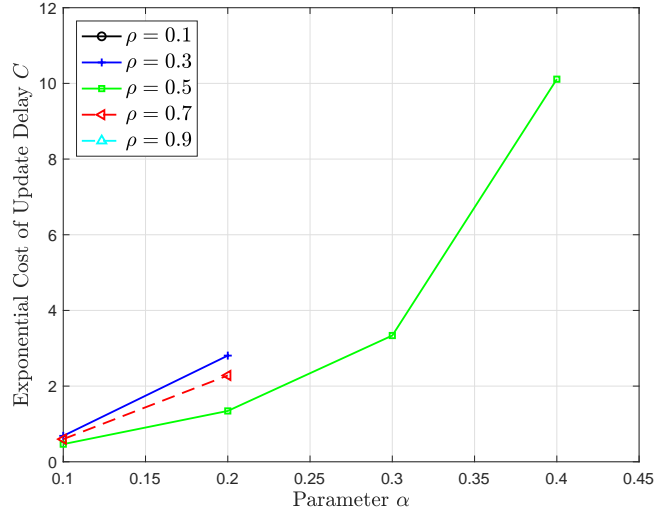}
	\vspace{-3mm}
	\caption{Average CoUD vs. the parameter $\alpha$ for the M/M/1 system with $\mu=1$, $\alpha<\lambda$, and $\alpha<\mu-\lambda$, exponential case.}
	\label{fig:exp_CoUD_vs_alpha}
\end{figure}

\begin{figure}[t!]\centering
	\centering
	\includegraphics[draft=false,scale=.5]{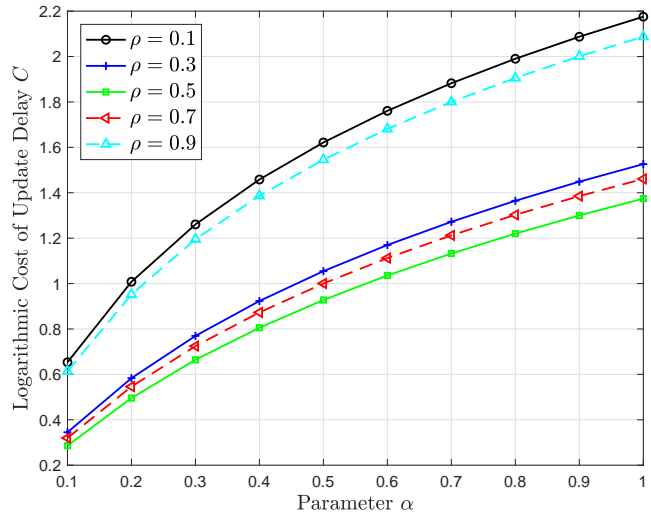}
	\vspace{-3mm}
	\caption{Comparison of the average CoUD vs. the parameter $\alpha$ for the M/M/1 system with $\mu=1$, logarithmic case.}
	\label{fig:log_CoUD_vs_alpha}
\end{figure}

In Fig.~\ref{fig:lin_CoUD_vs_alpha}-\ref{fig:log_CoUD_vs_alpha}, the average CoUD is shown as a function of the tuning parameter $\alpha$ for the three $f_s$ cases and different server utilizations $\rho$, for $\mu=1$.
Recall that for $f_s(t)=e^{\alpha t}-1$, we require that $\alpha<\lambda$ and $\alpha<\mu-\lambda$ as indicated in Theorem 2.
We observe that the linear cost function increases linearly with $\alpha$, the exponential cost function increases exponentially with $\alpha$, and equivalently, the logarithmic cost function increases logarithmically with $\alpha$.
This indicates how a change in the value of $\alpha$ will affect differently the three cases.

\section{Summary And Future Directions}
\label{sec:conclusions}
In this study, we have considered the characterization of the information transmitted over a source-destination link, modelled as an M/M/1 queue.
To capture freshness, we introduce the CoUD metric through three cost functions that can be chosen in relation with the autocorrelation of the process under observation.
To characterize the importance of an update, we define VoIU that measures the reduction of CoUD and therefore of uncertainty.
Either can be used depending on the application.
%CoUD and VoIU can be used interchangeably depending on the application.
We  analysed the relation between CoUD and VoIU and observed that convex and concave CoUD functions lead to a tradeoff between CoUD and VoIU, while linearity reflects only on the CoUD.
Moreover, we derived exact expressions, upper bounds in relation with PCoUD, and the optimal policies, in various settings.

Depending on the application we can choose the utilization that has as an objective either the minimization of CoUD or the maximization of VoIU.
A key in the flexibility of these notions is the potential for usage of non-linear functions
to represent them, giving ground to establish differentiated service
classes in monitoring systems.
In the linear CoUD case, VoIU is independent of the cost assigned per time unit.
In the exponential and logarithmic cases however, there is a tradeoff between CoUD and VoIU.
That is, the smaller the average CoUD, the smaller the average VoIU. 
%For instance, higher correlation among the samples decreases their value of information 
For high correlation among the samples, choosing $f_s(t)= \log(\alpha t+1)$ decreases their value of information and equivalently choosing $f_s(t)=e^{\alpha t}-1$ in low correlation has the opposite effect.
In the future, we will extend this work to capture the association  of specific source structures with the different cost functions.

\appendices

\section{Proof of Theorem \ref{theorem1}}\label{Appendix_A'}
To compute the average CoUD for the linear $f_s(\cdot)$ case we utilize \eqref{eq:av_Delta} and \eqref{eq:Qi_at}.
Hence, the terms $\mathbb{E}[Y^2]$ and $\mathbb{E}[YT]$ need to be calculated. We know that $Y$ is exponentially distributed with average arrival rate $\lambda$, so we have $\mathbb{E}[Y^2] = 2/\lambda^2$. For $\mathbb{E}[YT]$, consider that the system time of update $i$ is
\begin{equation}
T_i = W_i + S_i,
\label{eq:sysTime}
\end{equation}
where $W_i$ is the waiting time and $S_i$ is the service time of update $i$.
Since, the service time $S_i$ is independent of the $i$th interarrival time $Y_i$, we can write
\begin{equation}
\mathbb{E}[T_i Y_i] = \mathbb{E}[(W_i+S_i)Y_i] = \mathbb{E}[W_i Y_i]+\mathbb{E}[S_i]\mathbb{E}[Y_i],
\label{eq:expTiYi}
\end{equation}
where $\mathbb{E}[S_i] = 1/ \mu$ and $\mathbb{E}[Y_i] = 1/ \lambda$.
Moreover, we can express the waiting time of update $i$ as the remaining system time of the previous update minus the elapsed time between the generation of updates $(i-1)$ and $i$, i.e.,
\begin{equation}
W_i = (T_{i-1} - Y_i)^+.
\label{eq:W_i}
\end{equation}
Note that if the queue is empty then $W_i = 0$.
%and if packet $i-1$ is still in the system then $T_{i-1}$ is independent of $Y_i$.
Also note that when the system reaches steady state the system times are stochastically identical, i.e., $T =^{st} T_{i-1} =^{st} T_i$.
%Additionally, the probability density function (pdf) of the system time $T$ for the M/M/1 is \cite{Papoulis}
Thus, the conditional expectation of the waiting time $W_i$ given $Y_i=y$ can be obtained as
\begin{align}
\mathbb{E}[W_i|Y_i=y] & = \mathbb{E}[(T_{i-1}-y)^+|Y_i=y] = \mathbb{E}[(T-y)^+] \notag \\
& = \int_y^\infty (t-y) f_T(t)\:\mathrm{d}t = \frac{e^{-\mu(1-\rho)y}}{\mu(1-\rho)}.
\label{eq:Wi_cond_Yi}
\end{align}
The expectation $\mathbb{E}[W_i Y_i]$ is then obtained as
\begin{equation}
\mathbb{E}[W_i Y_i] =  \int_0^\infty y\: \mathbb{E}[W_i | Y_i=y] f_{Y}(y) \:\mathrm{d}y = \frac{\rho}{\mu^2(1-\rho)}.
\label{eq:expWiYi}
\end{equation}
Utilizing \eqref{eq:av_Delta}, \eqref{eq:Qi_at}, \eqref{eq:expTiYi}, and
\eqref{eq:expWiYi}, yields the average CoUD in \eqref{eq:av_C_lin}.

To compute the average PCoUD for the linear $f_s(\cdot)$ case we utilize \eqref{eq:A_i}.
We know that $Y$ is exponentially distributed with average arrival rate $\lambda$, so we have $\mathbb{E}[Y] = 1/\lambda$. 
Moreover, we know that $T$ is exponentially distributed with parameter $(\mu-\lambda)$, so we have $\mathbb{E}[T]=1/(\mu-\lambda)$. 
Then, the average PCoUD in \eqref{eq:av_A_lin} follows. 

Alternatively, the conditional expectation of the PCoUD $A_i$ given $Y_i=y$ can be obtained as
\begin{align}
\mathbb{E}[A_i|Y_i=y] & = \mathbb{E}[\alpha (T_{i}+y)|Y_i=y] = \mathbb{E}[\alpha (T+y)] \notag \\
& = \int_0^\infty \alpha (t+y) f_T(t)\:\mathrm{d}t = \alpha \left(  \frac{1}{\mu-\lambda}+y \right).
\label{eq:Ai_cond_Yi}
\end{align}
Then, the expectation $\mathbb{E}[A_i]$ is obtained as
\begin{equation}
\mathbb{E}[A_i] =  \int_0^\infty \: \mathbb{E}[A_i | Y_i=y] f_{Y}(y) \:\mathrm{d}y = \alpha \left( \frac{1}{\lambda}+\frac{1}{\mu-\lambda} \right).
\label{eq:expAi}
\end{equation}

\section{Proof of Theorem \ref{theorem2}}\label{Appendix_B'}

The Laplace transform of AoI for the M/M/1 system with an FCFS queue discipline is given by \cite{Inoue18_arXiv}
\begin{equation}
	C^{*}(s) = \frac{\lambda (\mu-\lambda) \left(\lambda s+(\mu+s)^2\right)}{(\lambda+s) (\mu+s)^2 (s+\mu-\lambda)}.
	\label{eq:C_Laplace}
\end{equation}

To obtain the pdf of CoUD for the M/M/1 system with an FCFS queue discipline we take the inverse Laplace transform of \eqref{eq:C_Laplace} and we have that 
%\begin{align}
%	C(t) &= .
%	\label{eq:C_pdf}
%\end{align}
\begin{equation}
	C(t) = \left( \frac{ \lambda^2 }{\lambda-\mu}-\mu \right) e^{-\mu t} +\frac{ \lambda \mu}{\mu-\lambda} e^{-\lambda t} 
	-\lambda \mu t e^{-\mu t} +\mu(1-\rho) e^{-\mu(1- \rho)t}.
	\label{eq:C_pdf}
\end{equation}
Then, the average CoUD is given by 
\begin{equation}
C_E = \int_{0}^{\infty} (e^{\alpha t} -1) C(t) \:\mathrm{d}t.
\end{equation}

In the stationary FCFS M/GI/1 queue the Laplace transform of PAoI is given by \cite[Lemma~24]{Inoue18_arXiv}
\begin{equation}
A^{*}(s) =  T^{*}(s) S^{*}(s) - T^{*}(s+\lambda) \frac{s S^{*}(s)}{s +\lambda}.
\label{eq:A_Laplace_MGI1}
\end{equation}
Then, the Laplace transform of PCoUD for the M/M/1 system with an FCFS queue discipline can be obtained as
\begin{equation}
	A^{*}(s) = \frac{\lambda \mu (\mu-\lambda) (\mu+2 s)}{(\lambda+s) (\mu+s)^2 (s+\mu-\lambda)}.
	\label{eq:A_Laplace}
\end{equation}

To obtain the pdf of PCoUD for the M/M/1 system with an FCFS queue discipline we take the inverse Laplace transform of \eqref{eq:A_Laplace} and we have that 
%\begin{equation}
%	A(t) = .
%	\label{eq:A_pdf}
%\end{equation}
\begin{equation}
A(t) = \frac{\mu (2 \lambda^2 -2 \lambda \mu+\mu^2) }{\lambda (\lambda-\mu)} e^{-\mu t} +\frac{\mu(\mu-\lambda)}{\lambda} e^{-(\mu-\lambda)t} +\frac{\lambda \mu}{\mu-\lambda} e^{-\lambda t} -\mu^2 t e^{-\mu t}.
\label{eq:A_pdf}
\end{equation}

Then, the average PCoUD is given by 
\begin{equation}
A_E = \int_{0}^{\infty} (e^{\alpha t} -1) A(t) \:\mathrm{d}t.
\end{equation}

\section{Proof of Corollary \ref{corollary3}}\label{Appendix_C'}
To compute the average CoUD bound for the exponential $f_s(\cdot)$ case we utilize \eqref{eq:av_Delta} and \eqref{eq:Qi_exp(t)}.
Hence, the terms $\mathbb{E}\left[e^{\alpha (Y+T)} \right]$,   $\mathbb{E}\left[e^{\alpha  T} \right]$, and $\mathbb{E}[Y]$, need to be calculated. Since $Y$ is exponentially distributed with average arrival rate $\lambda$, we have $\mathbb{E}[Y] = 1/\lambda$. 
Let us consider a random variable $\bar{Y}$ that is i.i.d. with $Y$ and independent of $T$.
Then, for the expected value $\mathbb{E}[Q]$ we obtain the terms
\begin{align}
\mathbb{E}\left[ e^{\alpha  T} \right] =\begin{cases} 
\frac{-\mu(1-\rho)}{\alpha -\mu(1-\rho)}&\mbox{, if } \alpha < \mu (1-\rho), \\
+\infty&\mbox{, otherwise},
\end{cases} 
\label{eq:expectation_Veat}
\end{align}
\begin{align}
\mathbb{E}\left[e^{\alpha (\bar{Y}+T)} \right] &= \int_0^\infty \mathbb{E}\left[e^{\alpha (\bar{Y}+T)} \big| \bar{Y}=y\right] f_Y(y) \:\mathrm{d}y = \notag \\
&=\begin{cases} 
\frac{\mu(1-\rho)\lambda}{[\alpha -\mu(1-\rho)](\alpha-\lambda)}&\mbox{, if } \alpha < \lambda, \\
+\infty&\mbox{, otherwise},
\end{cases} 
\label{eq:expectation_Veat2}
\end{align}
where 
\begin{equation}
	\mathbb{E}\left[ e^{\alpha (\bar{Y}+T)} \big| \bar{Y}=y \right] =\begin{cases} 
		\frac{-\mu(1-\rho)}{\alpha -\mu(1-\rho)} \: e^{a y}&\mbox{, if } \alpha <\mu (1-\rho), \\
		+\infty&\mbox{, otherwise}.
	\end{cases} 
\end{equation}
Using the fact that $Cov[Y,T] \leq 0$, one can show that $\mathbb{E}[e^{\alpha (Y+T)}] \leq \mathbb{E}[e^{\alpha (\bar{Y}+T)}]$.
After applying all the relevant expressions to \eqref{eq:av_Delta}, we find the average CoUD bound in \eqref{eq:av_C_exp}.

Furthermore, we consider the Laplace transform of the interarrival times and system times \cite{Kleinrock}
\begin{equation}
Y^{*}(s) \doteq  \int_0^\infty \: e^{-s t} f_{Y}(t) \:\mathrm{d}t =   \frac{\lambda}{s+\lambda},
\label{eq:Y_laplace}
\end{equation}
\begin{equation}
T^{*}(s) \doteq  \int_0^\infty \: e^{-s t} f_{T}(t) \:\mathrm{d}t =   \frac{\mu(1-\rho)}{s+\mu(1-\rho)}.
\label{eq:T_laplace}
\end{equation}
to obtain the Laplace of the PCoUD $A=\alpha(\bar{Y}+T)$ as
\begin{equation}
A^{*}(s) =  \frac{\alpha \lambda (\lambda-\mu)}{(\lambda+s) (\lambda-\mu-s)}.
\label{eq:A_laplace}
\end{equation}
%The derivative of $A^{*}(s)$ evaluated at $s=0$ provides us with the result in \eqref{eq:av_A_lin} and \eqref{eq:expAi}.
Taking the inverse Laplace transform of \eqref{eq:A_laplace} yields
\begin{equation}
A(t) = \frac{\alpha \lambda (\lambda-\mu) e^{-\lambda t} \left(1-e^{2 \lambda t-\mu t}\right)}{2 \lambda-\mu}.
\label{eq:A_pdf2}
\end{equation}
Finally, using the distribution of PCoUD in \eqref{eq:A_pdf2} we have that 
\begin{equation}
\mathbb{E}\left[e^{\alpha (\bar{Y}+T)} -1\right] =  \int_0^\infty (e^{a (\bar{y}+t)}-1)  A(t) \mathrm{d}t = \frac{\alpha (\mu-\alpha)}{(\alpha-\lambda) (\alpha+\lambda-\mu)},
\label{eq:expectation_Aexpat}
\end{equation}
equals the upper bound in \eqref{eq:av_C_exp}.
This completes the proof.

\section{Proof of Corollary \ref{corollary4}}\label{Appendix_D'}
To compute the average CoUD bound for the logarithmic $f_s(\cdot)$ case we utilize \eqref{eq:av_Delta} and \eqref{eq:Qi_log2}.
Hence, the terms $\mathbb{E}\left[(\alpha (Y+T)+1) \log(\alpha (Y+T)+1)\right]$, $\mathbb{E}\left[(\alpha  T+1)\log(\alpha T +1) \right]$, and $\mathbb{E}[Y]$, need to be calculated. Since $Y$ is exponentially distributed with average arrival rate $\lambda$, we have $\mathbb{E}[Y] = 1/\lambda$. 
Let us consider a random variable $\bar{Y}$ that is i.i.d. with $Y$ and independent of $T$.
Then, the expected value $\mathbb{E}[Q]$ can be obtained by the following terms:

Starting from the second term of \eqref{eq:Qi_log2} we have
\begin{align}
\mathbb{E}\big[\log(\alpha T+1) \big] =  \int_0^\infty \log(\alpha t+1)  \mathbb{P}_T(t) \mathrm{d}t =  -e^{\mu(1-\rho) /\alpha}  Ei\Big[-\frac{ \mu(1-\rho)}{\alpha}\Big] \mbox{, for}\:\:\:\lambda<\mu,
\label{eq:expectation_Vlogat}
\end{align}
where  $Ei$ denotes the exponential integral defined in \eqref{eq:exp_integr}.
Moreover,
\begin{align}
&\mathbb{E}\big[\alpha T \log(\alpha T+1) \big] = \alpha \int_0^\infty t \log(\alpha t+1)  \mathbb{P}_T(t) \mathrm{d}t = \notag\\   
& =- \frac{1}{\mu(1-\rho)} \Big[(\alpha-\mu(1-\rho)) e^{\mu(1-\rho) /\alpha}
 \times Ei{\Big[}-\frac{ \mu(1-\rho)}{\alpha}\Big] -\alpha \Big] \mbox{, for}\:\:\:\lambda<\mu.
\label{eq:expectation_Vlogat2}
\end{align}
Finally, the first term of \eqref{eq:Qi_log2} is given by
\vspace{-2mm}
\begin{align}
&\mathbb{E}\big[(\alpha (\bar{Y}+T)+1)\log(\alpha (\bar{Y}+T)+1) \big] =\notag  \\
& = \int_0^\infty \mathbb{E}\left[(\alpha (\bar{Y}+T)+1)\log(\alpha (\bar{Y}+T)+1) \big| \bar{Y}=y\right] f_Y(y) \:\mathrm{d}y = 
 - \frac{\lambda}{\mu(1-\rho)}  \int_0^\infty \Big[\alpha e^{\mu(1-\rho) (1/ \alpha+y)}  \notag  \\
 & Ei\Big[-\frac{\mu(1-\rho)(\alpha y+1)}{\alpha}\Big] - (\alpha y \mu (1-\rho)+\alpha+\mu(1-\rho)) \log(\alpha y +1) - \alpha \Big] e^{-\lambda y} \mathrm{d}y.  
\label{eq:expectation_Vlogat5}
\end{align}
This can be separated into three parts i.e., 

\emph{1}.
\begin{align}
& - \frac{\lambda}{\mu(1-\rho)}  \int_0^\infty \alpha e^{\mu(1-\rho) (1/ \alpha+y)}  Ei\Big[-\frac{\mu(1-\rho)(\alpha y+1)}{\alpha}\Big]  e^{-\lambda y} \mathrm{d}y = \notag  \\
& = - \frac{\lambda}{\mu(1-\rho)} \alpha e^{\mu(1-\rho) / \alpha} \int_0^\infty e^{[\mu(1-\rho) -\lambda] y}  Ei\Big[-\mu(1-\rho) y -\frac{\mu(1-\rho)}{\alpha}\Big] \mathrm{d}y = \notag  \\
&= \frac{\lambda}{\mu(1-\rho)} \alpha e^{\mu(1-\rho) / \alpha} \frac{1}{\mu-2\lambda} \left( Ei\left[ -\frac{\mu(1-\rho)}{\alpha}\right] - e^{(2 \lambda-\mu) /\alpha}  Ei\left[ -\frac{\lambda}{\alpha} \right] \right).
\label{eq:expectation_Vlogat6}
\end{align}

\emph{2}.
\begin{align}
& \frac{\lambda}{\mu(1-\rho)} \int_0^\infty (\alpha y \mu (1-\rho) +\alpha+\mu (1-\rho)) \log(ay+1) e^{-\lambda y} \mathrm{d}y = \notag \\
& = \lambda \alpha \int_0^\infty y \log(ay+1) e^{-\lambda y} \mathrm{d}y + \frac{\lambda (\alpha+\mu (1-\rho)) }{\mu(1-\rho)} \int_0^\infty \log(ay+1) e^{-\lambda y} \mathrm{d}y.
\label{eq:expectation_Vlogat7}
\end{align}

\emph{2a}.
\begin{equation}
\lambda \alpha \int_0^\infty y \log(ay+1) e^{-\lambda y} \mathrm{d}y = - \frac{1}{\lambda} \left[ (\alpha-\lambda) e^{\lambda/\alpha} Ei\left[- \frac{\lambda}{\alpha}\right] - \alpha \right]. 
\label{eq:expectation_Vlogat7a}
\end{equation}

\emph{2b}.
\begin{equation}
\frac{\lambda (\alpha+\mu (1-\rho)) }{\mu(1-\rho)} \int_0^\infty \log(ay+1) e^{-\lambda y} \mathrm{d}y = -\frac{1}{\mu(1-\rho)} (\alpha+\mu(1-\rho)) e^{\lambda/\alpha} Ei\left[- \frac{\lambda}{\alpha} \right]. 
\label{eq:expectation_Vlogat7b}
\end{equation}

\emph{3}.
\begin{align}
\frac{\lambda}{\mu(1-\rho)} \int_0^\infty \alpha e^{-\lambda y} =\frac{1}{\mu(1-\rho)} \alpha.
\label{eq:expectation_Vlogat8}
\end{align}
After applying all the relevant expressions to \eqref{eq:av_Delta}, we find the average CoUD bound in \eqref{eq:av_C_log_bound}.

Using the fact that $Cov[Y,T] \leq 0$, one can show that $\mathbb{E}[\log(\alpha (Y+T)+1)] \geq \mathbb{E}[ \log(\alpha (\bar{Y}+T)+1)]$.
Moreover, using the distribution of PCoUD in \eqref{eq:A_pdf2} we have that 
\begin{align}
\mathbb{E}\left[\log(\alpha (\bar{Y}+T)+1)\right] &=  \int_0^\infty \log(a (\bar{y}+t)+1)  A(t) \mathrm{d}t =\notag \\
&= \frac{e^{-\frac{\lambda}{\alpha}} \left(e^{\frac{2 \lambda}{\alpha}} (\lambda-\mu) Ei \left[\frac{\lambda}{\alpha}\right]+\lambda e^{\mu/\alpha} Ei \left[\frac{\mu-\lambda}{\alpha}\right]\right)}{2 \lambda-\mu}, \quad \lambda \neq \mu/2.
\label{eq:expectation_Alogat}
\end{align}
The limit of \eqref{eq:expectation_Alogat} as $\lambda$ approaches $\mu/2$ is $\frac{2 \alpha-e^{\frac{\mu}{2 \alpha}} (2 \alpha-\mu) Ei \left(-\frac{\mu}{2 \alpha}\right)}{\alpha \mu}$.
The expression in \eqref{eq:expectation_Alogat} equals the upper bound in \eqref{eq:av_C_log_bound}.
This completes the proof.

\section{Proof of Theorem \ref{theorem4}}\label{Appendix_E'}

For the $f_s(t)=\alpha t$ case, the expected value $\mathbb{E}[V]$ conditioned on the interarrival time $Y=y$ can be obtained as
\begin{align}
& \mathbb{E}\left[ \frac{Y}{Y+T} \Big| Y=y \right] = \mathbb{E}\left[ \frac{y}{y+T} \right] 
= \int_0^\infty \frac{y}{y+t}  \:\mathbb{P}_T(t) \mathrm{d}t =\notag \\ 
&= -y\mu(1-\rho) e^{y\mu(1-\rho)} Ei(-\mu(1-\rho)y)),
%& \quad \quad \quad \quad \times Ei(-\mu(1-\rho)y)) \mbox{, for}\:\:\:(\mu-\lambda)>0. 
\label{eq:expectation_Vat}
\end{align}
for $(\mu-\lambda)>0$.

Furthermore, using the iterated expectation and the probability density function of $Y$, \eqref{eq:expectation_Vat} implies
\begin{align}
\mathbb{E}\big[V_P\big] = \frac{(1-\rho)}{2\rho} \: {}_2 F_1 \left(1,2{;}3{;}2-\frac{1 }{\rho}\right),
\label{eq:av_E_V_lin}
\end{align}
where the integral is calculated with the help of \cite[6.228]{table} and ${}_2 F_1$ is the hypergeometric function.
Applying the obtained expression to \eqref{eq:av_V}, we find the average VoIU in \eqref{eq:av_V_lin}.

%\newpage

\bibliography{references}

\begin{thebibliography}{10}
\providecommand{\url}[1]{#1}
\csname url@samestyle\endcsname
\providecommand{\newblock}{\relax}
\providecommand{\bibinfo}[2]{#2}
\providecommand{\BIBentrySTDinterwordspacing}{\spaceskip=0pt\relax}
\providecommand{\BIBentryALTinterwordstretchfactor}{4}
\providecommand{\BIBentryALTinterwordspacing}{\spaceskip=\fontdimen2\font plus
\BIBentryALTinterwordstretchfactor\fontdimen3\font minus
  \fontdimen4\font\relax}
\providecommand{\BIBforeignlanguage}[2]{{%
\expandafter\ifx\csname l@#1\endcsname\relax
\typeout{** WARNING: IEEEtran.bst: No hyphenation pattern has been}%
\typeout{** loaded for the language `#1'. Using the pattern for}%
\typeout{** the default language instead.}%
\else
\language=\csname l@#1\endcsname
\fi
#2}}
\providecommand{\BIBdecl}{\relax}
\BIBdecl

\bibitem{Kosta17_ISIT}
A.~Kosta, N.~Pappas, A.~Ephremides, and V.~Angelakis, ``Age and value of
  information: Non-linear age case,'' in \emph{Proc. IEEE ISIT}, Jun. 2017, pp.
  326--330.

\bibitem{Kaul12_INFOCOM}
S.~Kaul, R.~Yates, and M.~Gruteser, ``Real-time status: How often should one
  update?'' in \emph{Proc. IEEE INFOCOM}, Mar. 2012, pp. 2731--2735.

\bibitem{Song1990}
X.~{Song} and J.~W.~S. {Liu}, ``Performance of multiversion concurrency control
  algorithms in maintaining temporal consistency,'' in \emph{Proc. IEEE 14th
  Annual COMPSAC}, Oct. 1990, pp. 132--139.

\bibitem{Segev1991}
A.~Segev and W.~Fang, ``Optimal update policies for distributed materialized
  views,'' \emph{Management Science}, vol.~37, no.~7, pp. 851--870, 1991.

\bibitem{Adelberg1995_ACM}
B.~Adelberg, H.~Garcia-Molina, and B.~Kao, ``Applying update streams in a soft
  real-time database system,'' in \emph{ACM SIGMOD Record}, vol.~24, no.~2,
  1995, pp. 245--256.

\bibitem{Cho2000_ACM}
J.~Cho and H.~Garcia-Molina, ``Synchronizing a database to improve freshness,''
  in \emph{ACM sigmod record}, vol.~29, no.~2, 2000, pp. 117--128.

\bibitem{Talak18_arXiv}
R.~Talak, S.~Karaman, and E.~Modiano, ``Can determinacy minimize age of
  information?'' \emph{arXiv:1810.04371}, 2018.

\bibitem{Inoue18_arXiv}
Y.~Inoue, H.~Masuyama, T.~Takine, and T.~Tanaka, ``A general formula for the
  stationary distribution of the age of information and its application to
  single-server queues,'' \emph{arXiv:1804.06139}, 2018.

\bibitem{Kam16}
C.~Kam, S.~Kompella, G.~D. Nguyen, and A.~Ephremides, ``Effect of message
  transmission path diversity on status age,'' \emph{IEEE Transactions on
  Information Theory}, vol.~62, no.~3, pp. 1360--1374, Mar. 2016.

\bibitem{Sun18_INFOCOM}
Y.~Sun, E.~Uysal-Biyikoglu, and S.~Kompella, ``Age-optimal updates of multiple
  information flows,'' in \emph{Proc. IEEE INFOCOM Workshops}, 2018, pp.
  136--141.

\bibitem{Najm18_INFOCOM}
E.~Najm and E.~Telatar, ``Status updates in a multi-stream {M/G/1/1} preemptive
  queue,'' in \emph{Proc. IEEE INFOCOM Workshops}, 2018, pp. 124--129.

\bibitem{Yates19_transactions}
R.~D. Yates and S.~K. Kaul, ``The age of information: Real-time status updating
  by multiple sources,'' \emph{IEEE Transactions on Information Theory},
  vol.~65, no.~3, pp. 1807--1827, 2019.

\bibitem{Costa16}
M.~Costa, M.~Codreanu, and A.~Ephremides, ``On the age of information in status
  update systems with packet management,'' \emph{IEEE Transactions on
  Information Theory}, vol.~62, no.~4, pp. 1897--1910, Apr. 2016.

\bibitem{Modiano15_ISIT}
L.~Huang and E.~Modiano, ``Optimizing age-of-information in a multi-class
  queueing system,'' in \emph{Proc. IEEE ISIT}, Jun. 2015, pp. 1681--1685.

\bibitem{Kaul12_CISS}
S.~K. Kaul, R.~D. Yates, and M.~Gruteser, ``Status updates through queues,'' in
  \emph{Proc. IEEE 46th Annual CISS}, Mar. 2012, pp. 1--6.

\bibitem{Najm16_ISIT}
E.~Najm and R.~Nasser, ``Age of information: The gamma awakening,'' in
  \emph{Proc. IEEE ISIT}, Jul. 2016, pp. 2574--2578.

\bibitem{Pappas15_ICC}
N.~Pappas, J.~Gunnarsson, L.~Kratz, M.~Kountouris, and V.~Angelakis, ``Age of
  information of multiple sources with queue management,'' in \emph{Proc. IEEE
  ICC}, Jun. 2015, pp. 5935--5940.

\bibitem{Kosta19_ISIT}
A.~Kosta, N.~Pappas, A.~Ephremides, and V.~Angelakis, ``Queue management for
  age sensitive status updates,'' in \emph{Proc. IEEE ISIT}, Jul. 2019, pp.
  1--5.

\bibitem{Kosta19_JCN}
A.~Kosta, N.~Pappas, A.~Ephremides, and V.~Angelakis, ``Age of information
  performance of multiaccess strategies with packet management,''
  \emph{IEEE/KICS Journal of Communications and Networks}, vol.~21, no.~3, pp.
  244--255, 2019.

\bibitem{Soysal19_arXiv}
A.~Soysal and S.~Ulukus, ``Age of information in {G/G/1/1} systems: Age
  expressions, bounds, special cases, and optimization,''
  \emph{arXiv:1905.13743}, 2019.

\bibitem{Kam2018_transactions}
C.~Kam, S.~Kompella, G.~D. Nguyen, J.~E. Wieselthier, and A.~Ephremides, ``On
  the age of information with packet deadlines,'' \emph{IEEE Transactions on
  Information Theory}, vol.~64, no.~9, pp. 6419--6428, Sept. 2018.

\bibitem{Baknina18_ISIT}
A.~Baknina, S.~Ulukus, O.~Oze, J.~Yang, and A.~Yener, ``Sening information
  through status updates,'' in \emph{Proc. IEEE ISIT}, Jun. 2018, pp.
  2271--2275.

\bibitem{Bedewy16_ISIT}
A.~M. Bedewy, Y.~Sun, and N.~B. Shroff, ``Optimizing data freshness,
  throughput, and delay in multi-server information-update systems,'' in
  \emph{Proc. IEEE ISIT}, Jul. 2016, pp. 2569--2573.

\bibitem{Bedewy17_ISIT}
A.~M. Bedewy, Y.~Sun, and N.~B. Shroff, ``Age-optimal information updates in
  multihop networks,'' in \emph{Proc. IEEE ISIT}, Jun. 2017, pp. 576--580.

\bibitem{Talak17_Allerton}
R.~Talak, S.~Karaman, and E.~Modiano, ``Minimizing age-of-information in
  multi-hop wireless networks,'' in \emph{Proc. IEEE 55th Annual Allerton},
  2017, pp. 486--493.

\bibitem{Yates18_arXivSHS}
R.~D. Yates, ``The age of information in networks: Moments, distributions, and
  sampling,'' \emph{arXiv:1806.03487}, 2018.

\bibitem{Yates18_INFOCOM}
R.~D. Yates, ``Age of information in a network of preemptive servers,'' in
  \emph{Proc. IEEE INFOCOM Workshops}, April 2018, pp. 118--123.

\bibitem{Kosta18_GLOBECOM}
A.~Kosta, N.~Pappas, A.~Ephremides, and V.~Angelakis, ``Age of information and
  throughput in a shared access network with heterogeneous traffic,'' in
  \emph{Proc. IEEE GLOBECOM}, Dec. 2018.

\bibitem{Stamatakis19_GLOBECOM}
G.~Stamatakis, N.~Pappas, and A.~Traganitis, ``Controlling status updates in a
  wireless system with heterogeneous traffic and {AoI} constraints,'' in
  \emph{Proc. IEEE GLOBECOM}, Dec. 2019.

\bibitem{Buyukates19_arXiv}
B.~Buyukates, A.~Soysal, and S.~Ulukus, ``Age of information in multicast
  networks with multiple update streams,'' \emph{arXiv:1904.11481}, 2019.

\bibitem{Yates15_ISIT}
R.~D. Yates, ``Lazy is timely: Status updates by an energy harvesting source,''
  in \emph{Proc. IEEE ISIT}, Jun. 2015, pp. 3008--3012.

\bibitem{Sun16_INFOCOM}
Y.~Sun, E.~Uysal-Biyikoglu, R.~Yates, C.~E. Koksal, and N.~B. Shroff, ``Update
  or wait: How to keep your data fresh,'' in \emph{Proc. IEEE INFOCOM}, Apr.
  2016, pp. 1--9.

\bibitem{Soleymani18_arXiv}
T.~Soleymani, J.~S. Baras, and K.~H. Johansson, ``Stochastic control with stale
  information--part i: Fully observable systems,'' \emph{arXiv:1810.10983},
  2018.

\bibitem{Stamatakis19_arXiv_GLOBECOM}
G.~Stamatakis, N.~Pappas, and A.~Traganitis, ``Control of status updates for
  energy harvesting devices that monitor processes with alarms,'' in
  \emph{Proc. IEEE GLOBECOM Workshops}, Dec. 2019.

\bibitem{Altman19_Mobihoc}
E.~Altman, R.~El-Azouzi, D.~S. Menasche, and Y.~Xu, ``Forever young: Aging
  control for hybrid networks,'' in \emph{Proc. 20th ACM Mobihoc}, Jul. 2019,
  pp. 91--100.

\bibitem{Azouzi12_ICST}
R.~El-Azouzi, D.~S. Menasche, Y.~Xu \emph{et~al.}, ``Optimal sensing policies
  for smartphones in hybrid networks: A pomdp approach,'' in \emph{Proc. 6th
  Int. ICST Conf. on Perf. Eval. Methodologies and Tools (VALUETOOLS)}, 2012,
  pp. 89--98.

\bibitem{Cho2003_ACM}
J.~Cho and H.~Garcia-Molina, ``Effective page refresh policies for web
  crawlers,'' \emph{ACM Transactions on Database Systems (TODS)}, vol.~28,
  no.~4, pp. 390--426, 2003.

\bibitem{Even2007_ACM}
A.~Even and G.~Shankaranarayanan, ``Utility-driven assessment of data
  quality,'' \emph{ACM SIGMIS Database: the DATABASE for Advances in
  Information Systems}, vol.~38, no.~2, pp. 75--93, 2007.

\bibitem{Heinrich2009_JDIQ}
B.~Heinrich, M.~Klier, and M.~Kaiser, ``A procedure to develop metrics for
  currency and its application in crm,'' \emph{Journal of Data and Information
  Quality (JDIQ)}, vol.~1, no.~1, p.~5, 2009.

\bibitem{Ioannidis09_INFOCOM}
S.~Ioannidis, A.~Chaintreau, and L.~Massoulie, ``Optimal and scalable
  distribution of content updates over a mobile social network,'' in
  \emph{Proc. IEEE INFOCOM}, 2009, pp. 1422--1430.

\bibitem{Razniewski2016_ACM}
S.~Razniewski, ``Optimizing update frequencies for decaying information,'' in
  \emph{Proc. 25th ACM Int. Conf. on Information and Knowledge Management},
  2016, pp. 1191--1200.

\bibitem{Sun2017_transactions}
Y.~Sun, E.~Uysal-Biyikoglu, R.~D. Yates, C.~E. Koksal, and N.~B. Shroff,
  ``Update or wait: How to keep your data fresh,'' \emph{IEEE Transactions on
  Information Theory}, vol.~63, no.~11, pp. 7492--7508, 2017.

\bibitem{Sun18_JCN}
Y.~Sun and B.~Cyr, ``Sampling for data freshness optimization: Non-linear age
  functions,'' \emph{IEEE/KICS Journal of Communications and Networks},
  vol.~21, no.~3, pp. 204--219, 2019.

\bibitem{Sun18_SPAWC}
Y.~Sun and B.~Cyr, ``Information aging through queues: A mutual information
  perspective,'' in \emph{Proc. IEEE SPAWC}, June 2018, pp. 1--5.

\bibitem{Poojary17_ITW}
S.~Poojary, S.~Bhambay, and P.~Parag, ``Real-time status updates for correlated
  source,'' in \emph{Proc. IEEE ITW}, 2017, pp. 274--278.

\bibitem{Sun17_ISIT}
Y.~Sun, Y.~Polyanskiy, and E.~Uysal-Biyikoglu, ``Remote estimation of the
  wiener process over a channel with random delay,'' in \emph{Proc. IEEE ISIT},
  Jun. 2017, pp. 321--325.

\bibitem{Kam18_SPAWC}
C.~Kam, S.~Kompella, G.~D. Nguyen, J.~E. Wieselthier, and A.~Ephremides,
  ``Towards an \lq\lq{}effective age\rq\rq{} concept,'' in \emph{Proc. IEEE
  SPAWC}, June 2018, pp. 1--5.

\bibitem{Kam18_INFOCOM}
C.~Kam, S.~Kompella, G.~D. Nguyen, J.~E. Wieselthier, and A.~Ephremides,
  ``Towards an effective age of information: Remote estimation of a markov
  source,'' in \emph{Proc. IEEE INFOCOM}, 2018, pp. 367--372.

\bibitem{He18_INFOCOM}
Q.~He, G.~D{\'a}n, and V.~Fodor, ``Minimizing age of correlated information for
  wireless camera networks,'' in \emph{Proc. IEEE INFOCOM}, April 2018, pp.
  547--552.

\bibitem{Hribar17_GLOBECOM}
J.~Hribar, M.~Costa, N.~Kaminski, and L.~A. DaSilva, ``Updating strategies in
  the internet of things by taking advantage of correlated sources,'' in
  \emph{Proc. IEEE GLOBECOM}, Dec 2017, pp. 1--6.

\bibitem{NET-060}
A.~Kosta, N.~Pappas, and V.~Angelakis, ``Age of information: A new concept,
  metric, and tool,'' \emph{Foundations and
  Trends$\textsuperscript{\textregistered}$ in Networking}, vol.~12, no.~3, pp.
  162--259, 2017.

\bibitem{Costa14_ISIT}
M.~Costa, M.~Codreanu, and A.~Ephremides, ``Age of information with packet
  management,'' in \emph{Proc. IEEE ISIT}, Jun. 2014, pp. 1583--1587.

\bibitem{Kleinrock}
L.~Kleinrock, \emph{{Queueing Systems}}.\hskip 1em plus 0.5em minus 0.4em\relax
  Wiley Interscience, 1975, vol. I: Theory.

\bibitem{table}
I.~S. Gradshteyn, I.~M. Ryzhik, A.~Jeffrey, and D.~Zwillinger, \emph{Table of
  integrals, series, and products}.\hskip 1em plus 0.5em minus 0.4em\relax
  Elsevier 2007.

\end{thebibliography}
\bibliographystyle{IEEEtran}

\end{document}